\documentclass[a4paper,twocolumn,superscriptaddress,11pt]{quantumarticle}

\usepackage{wrapfig}
\usepackage{mathtools}
\usepackage{amsfonts}
\usepackage{latexsym}
\usepackage{relsize}
\usepackage{mathrsfs}

\DeclareMathAlphabet{\mathbbold}{U}{bbold}{m}{n}

\usepackage{euscript}
\usepackage{amssymb}
\usepackage{graphicx}
\usepackage{amsmath}
\usepackage{amsbsy}
\usepackage[font=scriptsize]{caption}
\usepackage{subcaption}

\usepackage{amsthm}
\usepackage{float}

\usepackage{bbm}
\usepackage{bm}
\usepackage{epsfig}
\usepackage{epstopdf}
\usepackage{dsfont}
\usepackage{soul,ulem}

\usepackage[colorlinks]{hyperref}
\usepackage[figure,table]{hypcap}
\usepackage{enumerate}
\usepackage{geometry}
\usepackage{geometry}
\hypersetup{
	bookmarksnumbered,
	pdfstartview={FitH},
	citecolor={darkgreen},
	linkcolor={darkred},
       linktoc={page},
	urlcolor={darkblue},
	pdfpagemode={UseOutlines}}

\usepackage{color}

\definecolor{darkgreen}{RGB}{40,130,40}
\definecolor{darkblue}{RGB}{0,0,190}
\definecolor{darkred}{RGB}{238,0,0}
\usepackage{float}

\def\EQ#1{\begin{eqnarray}#1\end{eqnarray}}
\newcommand{\djj}{d\kern-0.4em\char"16\kern-0.1em}

\newcounter{lem}

\newtheorem{prop}{Proposition}\def\PRO{\begin{prop}}\def\ORP{\end{prop}}
\newtheorem{coro}{Corollary}\def\COR{\begin{coro}}\def\ROC{\end{coro}}
\newtheorem{theo}{Theorem}\def\TH{\begin{theo}}\def\HT{\end{theo}}
\def\TH{\begin{theo}}\def\HT{\end{theo}}
\newtheorem{defi}[prop]{Definition}\def\DE{\begin{defi}}\def\ED{\end{defi}}
\newtheorem{lemme}[lem]{Lemma}\def\LE{\begin{lemme}}\def\EL{\end{lemme}}

\def\ket#1{\left| #1 \right\rangle}

\def\dm#1{\left|#1 \right\rangle \left\langle #1 \right|}

\makeatletter
\newcommand{\shorteq}{%
  \settowidth{\@tempdima}{-}
  \resizebox{\@tempdima}{\height}{=}%
}
\makeatother

\def \beq {\begin{equation}}
\def \eeq {\end{equation}}
\def \ba {\begin{eqnarray}}
\def \ea {\end{eqnarray}}

\begin{document}

\title{Exponential improvements for quantum-accessible reinforcement learning}

\author{Vedran Dunjko}
\email{v.dunjko@liacs.leidenuniv.nl}
\affiliation{LIACS, Leiden University, Niels Bohrweg 1, 2333 CA Leiden, The Netherlands}
\affiliation{Max-Planck-Institut f\"{u}r Quantenoptik, Hans-Kopfermann-Str. 1, D-85748 Garching, Germany}
\author{Yi-Kai Liu}
\email{yi-kai.liu@nist.gov}
\affiliation{Joint Center for Quantum Information and Computer Science, University of Maryland, College Park, MD 20742, USA}
\affiliation{Applied and Computational Mathematics Division, National Institute of Standards and Technology, Gaithersburg, MD 20899, USA}
\author{Xingyao Wu}
\email{wu.x.yao@gmail.com}
\affiliation{Joint Center for Quantum Information and Computer Science, University of Maryland, College Park, MD 20742, USA}
\author{Jacob M. Taylor}
\email{jmtaylor@umd.edu}
\affiliation{Joint Center for Quantum Information and Computer Science, University of Maryland, College Park, MD 20742, USA}
\affiliation{Joint Quantum Institute, National Institute of Standards and Technology, Gaithersburg, MD 20899, USA} 
\affiliation{Research Center for Advanced Science and Technology, University of Tokyo, Meguro-ku, Tokyo 153-8904, Japan}

\begin{abstract}

Quantum computers can offer dramatic improvements over classical devices for data analysis tasks such as prediction and classification. However, less is known about the advantages that quantum computers may bring in the setting of reinforcement learning, where learning is achieved via interaction with a task environment. 
Here, we consider a special case of reinforcement learning, where the task environment allows quantum access. In addition, we impose certain ``naturalness'' conditions on the task environment, which rule out the kinds of oracle problems that are studied in quantum query complexity (and for which quantum speedups are well-known). 

Within this framework of quantum-accessible reinforcement learning environments, we demonstrate that quantum agents can achieve exponential improvements in learning efficiency, surpassing previous results that showed only quadratic improvements. 
A key step in the proof is to construct task environments that encode well-known oracle problems, such as Simon's problem and Recursive Fourier Sampling, while satisfying the above ``naturalness'' conditions for reinforcement learning. Our results suggest that quantum agents may perform well in certain game-playing scenarios, where the game has recursive structure, and the agent can learn by playing against itself.
\end{abstract}

\maketitle
\section{Introduction}

Quantum machine learning (QML) is a relatively new discipline that investigates the interplay between quantum information processing and machine learning (ML) \cite{2017_Biamonte,2018_Dunjko}. 
Thus far, most of the attention in QML has focused on speed-ups in data analysis settings, namely supervised learning (\textit{e.g.,} classification) and unsupervised learning (\textit{e.g.,} clustering) \cite{2013_Aimeur, 2015_Wiebe, 2014_Rebentrost, 2016_Lloyd, 2015_Zhao}. 
 
From a more foundational perspective, computational learning theory (COLT) has also been extended to the quantum setting, and various results regarding classical-quantum separations are known (see \cite{2017_deWolf} for a recent review).

\begin{figure} 
 \includegraphics[width=0.48\textwidth,clip=true,trim =140 186 200 170]{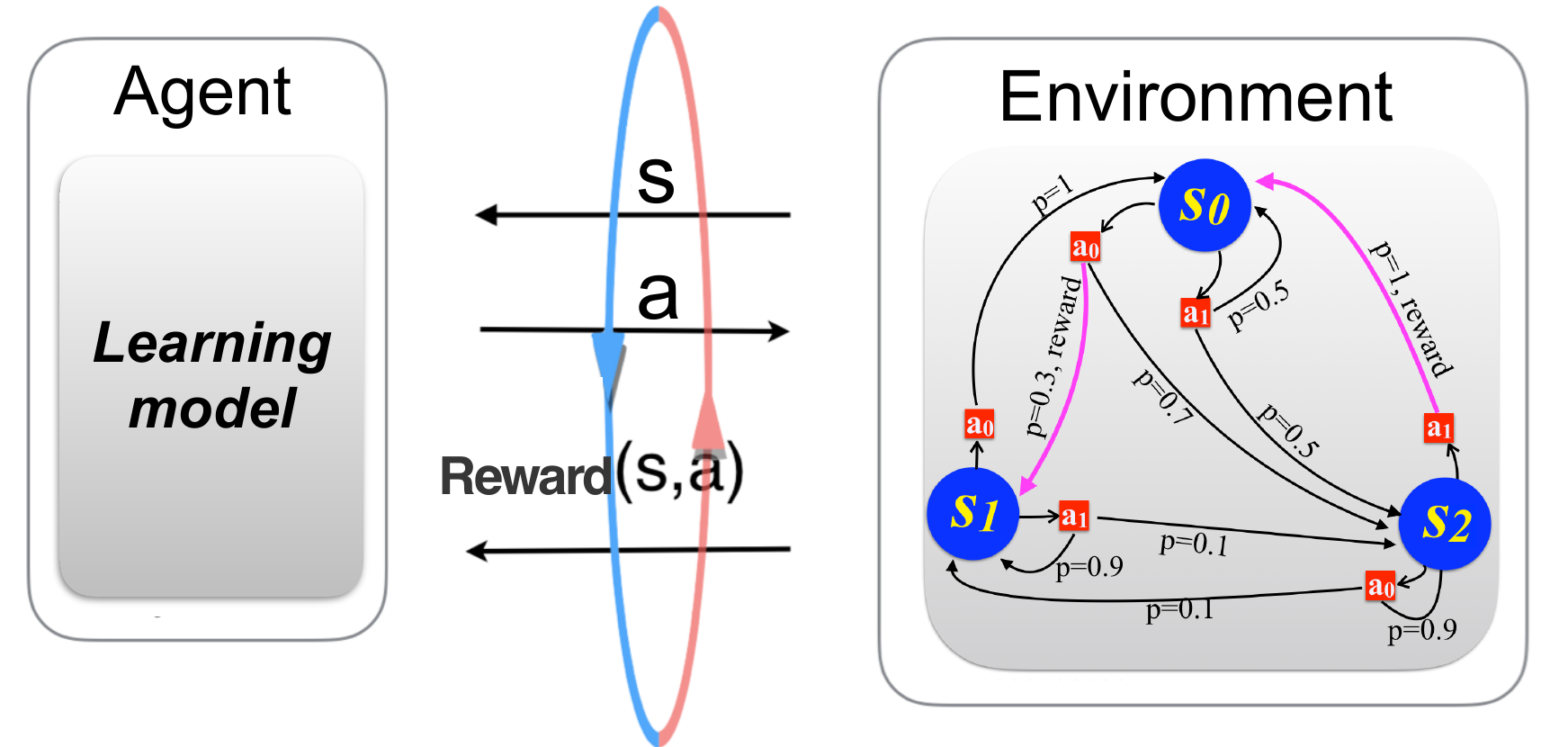}
\caption{\label{RL1} Illustration of the the basic agent-environment paradigm: an agent navigates a task environment (e.g. a maze) by taking actions and by receiving percepts (signals s, e.g. labels of possible positions in the maze). In RL, percepts can be rewarding. Basic environments are characterized by Markov decision processes (inset in environment), in which case the percepts are the states of the environment. }
 \end{figure}

Beyond supervised and unsupervised learning settings, which essentially deal with data analysis, \textit{reinforcement learning} (RL) \cite{1998_Sutton}, deals with interactive learning settings, and constitutes an effective bridge between data-analysis oriented ML and full-blown artificial intelligence (AI)  \cite{2009_Russel}. In RL we deal with a \textit{learning agent}, which learns by interacting with its \textit{task environment}.
In RL, the agent perceives (aspects of) the states of a task environment, and influences subsequent states by performing actions. Certain state-action-state transitions are rewarding, and successful learning agents learn \textit{optimal behavior} (see Fig~\ref{RL1} for an illustration).

RL is closely linked to robotics and AI tasks, and is thus also practically very well-motivated. For instance, RL plays a pivotal role in modern learning technologies -- from artificial personal assistants and self-driving cars, to the celebrated AlphaGo system which, startlingly, surpassed human-level gameplay in Go \cite{2016_Silver}. 
While this initial exciting result relied on various flavors of learning to achieve superior game-play, the most recent, and strongest variants (which now beat also best humans and software in chess), no longer utilize supervised learning from human examples and rely on RL and self-play \cite{2017_Silver,2017_Silver_b}.

\setlength{\intextsep}{1pt}%
\setlength{\columnsep}{10pt}%

Possible quantum enhancements in this more general learning setting have been explored only in few works. 
A few authors have considered scenarios where the internal computation of the agent is quantized, while the interaction with the environment remains classical.
For instance, in \cite{2014_Paparo}, a quantum algorithm that quadratically speeds up a variant of the Projective Simulation \cite{2012_Briegel} model was proposed. In  \cite{2016_Crawford} it was investigated whether quantum annealers could offer computational speed-ups for Boltzman machine-based RL engines. 

To go beyond purely ``internal speed-ups,'' other authors \cite{2015b_Dunjko,2016_Dunjko} 
have considered environments that are ``quantum-accessible,'' in the sense that they maintain superpositions, and allow exquisite quantum control. The relationship between purely classical environments and quantum accessible environments is analogous to
the relationship between classical and quantum oracles.

Given access to quantum-accessible environments, the agent-environment interaction (see Fig. \ref{RL1}) can also be also quantized, essentially as a conventional two-party quantum communication setting.
 In this framework, certain conditions were identified, which allow quadratic improvements in \textit{learning efficiency} (an analog of query complexity) for a class of RL scenarios, by utilizing amplitude amplification \cite{2015b_Dunjko,2016_Dunjko}. Certain criteria prohibiting improvements have been identified as well.  
In this work, we continue the investigation of \textit{the limits of speed-ups in learning efficiency}, given such quantum-accessible environments, and address the question of whether RL settings allow super-polynomial or exponential speed-ups, at least in certain cases.

The study of RL with quantum-accessible environments bears an obvious resemblance to the study of quantum query complexity, i.e., the study of quantum algorithms for oracle problems. However, RL has a different emphasis than query complexity. At a conceptual level, RL is more concerned with learning how to perform some task that involves the environment, such as playing a game; whereas query complexity is more concerned with characterizing some property of the oracle, such as whether there exists an input that causes the oracle to accept. 

In this paper, we state some simple conditions that separate RL problems from oracle problems. 
We then present examples of RL task environments 

where quantum-enhanced agents achieve optimal behaviours in polynomial time (in the size of the task environment), but where any classical learner requires exponential time to achieve equal levels of efficiency. 
These constructions are in fact based on oracle problems -- specifically, Simon's problem \cite{1995_Simon} and Recursive Fourier Sampling \cite{1997_Bernstein, 2008_Hallgren} -- but with suitable modifications that force a quantum agent to learn how to interact with the environment in a nontrivial way.

In order to achieve these provable quantum speed-ups, we consider RL task environments that may seem somewhat artificial and unrealistic, as these environments allow quantum access, and they encode rigid mathematical structures. However, we argue that these kinds of environments can actually occur in practical applications involving game-playing, such as the celebrated AlphaGo and AlphaGo Zero systems \cite{2016_Silver,2017_Silver,2017_Silver_b}. In these situations, an agent can simulate the game internally, and can learn by playing against itself. Hence, an agent with a quantum computer can simulate quantum access to an environment that encodes the game. Moreover, such games are built recursively from sub-games, in a way that is reminiscent of the Recursive Fourier Sampling problem and its generalizations \cite{1997_Bernstein, 2008_Hallgren}. Our results can be interpreted as further evidence that quantum agents can achieve super-polynomial improvements in learning to play these kinds of games.
This perspective on our results is reminiscent to results obtained for boolean formula evaluation tasks, such as in the study of NAND trees, which are, in fact game trees. Here, the levels of the tree correspond to alternating moves of two players, and the value of the node specifies whether the given player wins under perfect play. Previous works have shown polynomial speed-ups for generic game trees \cite{2008_Reichardt}, and even superpolynomial speed-ups for special families \cite{2012_Zhan}. The two perspectives are closely related, and there may be interest in combining the two approaches for the specific purposes of game play -- indeed, combining a heuristic for estimating values of nodes in game trees (Monte Carlo Tree Search), with reinforcement learning is at the basis of the strongest game playing results \cite{2017_Silver,2017_Silver_b}. However our overall objective pertains to general RL scenarios (characterized by Markov Decision Processes), where game playing is just one instance of possible applications.

The remainder of the paper is split into the section \ref{sec:tb} which covers the basic technical background, section \ref{sec:main} which discusses the embedding of oracle identification problems into RL tasks, and presents the criteria for ``genuine'', \textit{i.e.}, interactive RL problems and in section \ref{exp:sec}  we provide our main results. We finish off with the final discussion in section \ref{sec:dis}.

\section{Technical background}

\label{sec:tb}
To present the main results of our work in a self-contained manner, we first introduce the basic concepts from classical RL theory, from the quantum agent-environment and quantum RL framework introduced in  \cite{2016_Dunjko}, and from quantum oracle identification theory we will use later.

\subsection{The setting of reinforcement learning}
In reinforcement learning, an agent $A$ and an environment $E$ interact by the exchange of actions, from the set $\mathcal{A} = \{ a_i\}$, and percepts, from the set $\mathcal{S} = \{ s_i\}$. We consider finite sets of  actions and percepts. Further, in RL, the agent is driven to correct behavior by the issuing of rewards, from some ordered set $\Lambda$, \textit{e.g.} $\Lambda = \{0,1\},$ or $\Lambda \subseteq \mathbbmss{R}$. Basic environments are specified by \textit{Markov decision processes} (MDPs), characterized by the action, environmental state\footnote{In the context of MDP environments, the percepts are just the environmental states.} and reward sets, a stochastic \textit{transition function} $P_T(s_j | s_i, a_k),$ specifying the transition probability from state $s_i$ to $s_j$ under action $a_k$, and a  stochastic \textit{reward function} $R(s_i, a_j,s_k) \in Distr(\Lambda),$ which to each arc $(s_i,a_j,s_k)$ (probabilistically) assigns a reward. 
A \textit{policy} (of an agent) $\{ \pi(a|s)\}_s$  specifies the probability of (an agent outputting) an action $a$ given the state $s$.

Given an MDP $M$, we can identify various notions of optimal policies, \textit{e.g.} those which maximize the expected reward over some finite period $N$ (\textit{finite horizon}) or with respect to an \textit{$\gamma-$infinite horizon}. The latter is given with $R^{\infty}_\pi  = \lim_{l \rightarrow \infty}E_{\pi,M}[ \sum_{k=0}^{l} \gamma^k R^{\pi}_k],$ where $R_k$ is the reward  at the $k^{th}$ step (the geometrically decaying factor $\gamma^k$ ensures convergence, and increases the values of more immediate rewards). If we only care about the value of a policy after some number of initial steps $p$, we talk about efficiency after $p-$steps, given with $R^{\infty,p}_\pi  = \lim_{l \rightarrow \infty}E_{\pi,M}[ \sum_{k=p}^{l} \gamma^k R^{\pi}_k]$.
A learning agent $A$ is  \textit{$(\epsilon,\delta)-$efficient after $p$ steps} if the infinite horizon rewards $R$ of $A$ measured after the $k^{th}$ step for the agent $A$, satisfy $R^{\infty,k}_{\pi^\ast} \leq R + \epsilon,$  except with probability $\delta$. Analogous definition holds for the finite-horizon case. In other words, such agents are after $p$ steps (almost) as efficient as optimal agents.

 In other words, the agent is  $(\epsilon,\delta)-$efficient after $N$ steps, if after $N$ steps, except with probability $\delta$, it becomes equally (up to $\epsilon$) rewarded as an agent adhering to an optimal policy.
 
An environment (MDP) is:  \textit{episodic}, if the environment is re-set to (a set of) initial state(s)\footnote{Often we have the case that the last action of the agent results in feedback, provided by the first state of the next episode -- in this case we may deal with a set of initial states, which all have identical outbound transitions, \textit{i.e.,} the identity of the state we are in does not (directly) influence what happens in the current episode. } once a rewarding transition has occurred (if the rewards are stochastic, obtaining a reward of zero value still causes a re-set), and \textit{strictly $\eta-$episodic}, if the environment  is re-set to the (set of) initial state(s) after exactly $\eta$ steps.\vspace{0.1cm}
An environment has \textit{immediate rewards} if the occurrence of any state-action $(s,a)$ pair consistent with an optimal policy yields the largest (average) reward, i.e. for any other $a'$ the expected immediate reward of $(s,a')$ is smaller. In the simplest case of deterministic binary rewards, this simply means that every correct move is rewarded. Otherwise the setting has \textit{delayed rewards} -- a prototypical example is a maze problem, where rewards are issued only once the maze is solved, although there are obvious ``wrong'' and ``correct'' moves along the way.\vspace{0.2cm}
We should point out that not all environments correspond to MDPs: environments can also be partially observable (in which case the agent only perceives some noisy function of the environmental state), but in this work we will focus on fully observable settings~\footnote{It should be noted that quantum \textit{generalizations} of partially observable MDPs have been considered previously \cite{2014_Barry}. In the context of this work, however, we deal with environments which are specified by fully classical MDPs, but which are ``accessed'' in a quantum fashion, as explained shortly. }.

\subsection{Framework for quantum reinforcement learning}
\label{sec-oracul}
The generalization of the agent-environment setting is straightforward. The percept and action sets are promoted to sets of (also mutually) orthonormal kets $\{ \ket{s} | s \in \mathbf{S}\}$ and $\{ \ket{a} | a \in \mathbf{A} \}$.

The agent and the environment are modelled as (infinite) sequences of completely positive trace-preserving (CPTP) maps
$\{ \mathcal{M}_A^i\}_i$ and $\{ \mathcal{M}_E^i\}_i$, acting on the Hilbert spaces $H_A \otimes H_C$ and $H_C\otimes H_E$, respectively. Here, $H_A$, $H_C$ and $H_E$ specify the memory of the agent, the agent-environment interface (the \textit{communication channel}), and the memory of the environment, respectively.  The classical setting is recovered by restricting the agents and environments to classical maps, see \cite{2015b_Dunjko} for a formal definition, and Section \ref{SI} for further information.

Recall that, in the (fully observable) classical case, the task environment could be completely described by an MDP. However, if we allow quantum interaction, then the MDP no longer provides a complete description, because it does not specify the behavior of the environment on superpositions of actions and percepts. Indeed, there are many possible quantum environments that have identical behavior on classical inputs, and hence correspond to the same MDP. 

Each such environment we call \textit{a quantum realization of the MDP}.  

We are interested in quantum environments that preserve superpositions of actions and percepts. Intuitively, we might expect that such a ``nice'' quantum environment should always exist, because a mixed quantum state can always be purified, and a quantum channel can always be implemented by a unitary operation acting on a larger system. 

In fact, this claim can be made rigorous, as follows. In \cite{2015b_Dunjko} it was shown that for any $\eta-$episodic environment (which does not need to be fully observable), there exists a realization which is also \textit{quantum-accessible}.
This latter property implies that the agent can utilize its access to 
the environment to simulate an oracle $E_q$ that has the following behavior:
\EQ{
 \ket{a_1, \ldots, a_\eta}\ket{y} \stackrel{E_q}{\longrightarrow} \ket{a_1, \ldots, a_\eta}\ket{y \oplus R(a_1, \ldots, a_\eta)}  \nonumber,}
where $R(a_1, \ldots, a_\eta)$ is the reward value obtained by the agent once the agent executes the sequence of actions $a_1, \ldots, a_\eta$,\footnote{Note that this is a well-defined quantity only in deterministic environments, where the action sequence deterministically specifies the corresponding state sequence, and reward values.} and $\oplus$ denotes addition in the appropriate group. 

The process of simulating access to $E_q$ in a quantum-accessible environment is called \textit{oraculization}. 
Here, each query to the oracle $E_q$ requires approximately $5 \eta $ interaction steps with the environment. 
(More details are provided in Section \ref{SI}.) This allows us to utilize techniques from quantum algorithms (e.g., oracle identification) for reinforcement learning.

The basic idea of our overall approach can be summarized as follows. The learning of the the quantum-enhanced agent is split into two phases. In the first phase, it will utilize quantum interaction (via the \textit{oraculization} process) with the task environment, to effectively simulate access to a quantum oracle, which conceals critical information about the environment.  In the second phase, it will use this information to quickly find optimal behavior in the given task environment.

\subsection{Oracle identification and Simon's problem}

The first provable quantum improvements over classical computation involved the use of oracles, specifically the problems of oracle identification.  For our purposes,
 we specify the task as follows: for a (finite) set of oracles $O = \{ \mathcal{O}_i \}_i,$ which is partitioned as a collection of disjoint subsets ($O = \bigcup_{j} O_j,\ O_k \cap O_{k'} = \emptyset$, unless $k=k'$), given access to an oracle $\mathcal{O},$ identify which subset $O_j$ it belongs to. 
 In the cases we consider, the oracles will evaluate a boolean function, and the task will be to identify to which specified collection (out of exponentially many) of boolean functions the given instance belongs to\footnote{In more technical terms, we will be dealing with promise problems, where the collections do not comprise the entire set of possible functions. It is well-known that exponential separations in oracle identification, or computational learning, are not possible without such promises, see, \textit{e.g.},  \cite{BB01,2017_deWolf}.  }. 
 First examples here were the Deutsch and the Deutsch-Jozsa algorithm, where the set of oracles contained all boolean functions which are constant (subset of 2 elements) or balanced (subset of $\left({n \atop n/2} \right)$ elements), basic Grover's search \cite{1996_Grover} (where the subsets are singlet sets, and the oracles are promised attain value $1$ for one element and zero otherwise). Depending on the details of figures of merit one considers, and the exact specification of what oracles do, this framework also captures COLT as well \footnote{If the oracles are functions which can be queried, then this constitutes the standard oracular setting, or, similarly, learning from membership queries. However, they could also be objects which produce samples from unknown distributions, in which case we are broaching the conventional probably approximately correct learning. }. 

One of the first examples of oracle identification, where a strict exponential separation between classical and quantum computation was proven is, Simon's problem \cite{1995_Simon}.

In Simon's construction, we are given an oracle encoding a function $f: \{0,1\}^{n} \rightarrow \{0,1\}^n$, with the promise 
that there exists a secret string $s \in \{ 0,1\}^n$, such that $f(x) = f(y)$
 if and only if $x = y \oplus s$, where $\oplus$ is the element-wise modulo 2 addition. 
 Intuitively, to identify $s,$ one must find two strings which attain the same value under $f$, which, again intuitively, requires $O(2^{n})$ steps. A more careful analysis proves that any algorithm which finds $s$ with non-negligible probability needs $\Omega(2^{n/2})$ queries \cite{1995_Simon}. Access to a quantum oracle mapping $\ket{x}\ket{y} \mapsto \ket{x} \ket{y \oplus f(x)}$, allows an efficient quantum algorithm for finding $s$, and this can be done, e.g., by using a probabilistic algorithm \cite{1995_Simon} using $O(n)$ queries (which implies a zero-error algorithm with expected polynomial running time), or a zero-error deterministic algorithm with polynomial worst-case running time \cite{1997_Brassard}.

We point out that some oracle identification problems such as Fourier sampling (discussed in \ref{SI}) and Simon's problem play a role in the early results of quantum COLT: in the seminal work of \cite{1998_Bshouty} the quantum Fourier sampling algorithm is used to efficiently learn DNF formulas just from examples under the uniform distribution, and in \cite{2004_Servedio}, Simon's algorithm is used to prove that if one-way functions exist, then there is a superpolynomial gap between classical and quantum exact learnability. In this work, the use of oracular separations for RFS and Simon's problem are, however, simpler and less subtle than in these quantum COLT works.

\label{sect:Simon}

\subsection{Trivial transformations of oracle problems into MDPs}

The framework of reinforcement learning is very broad, and many kinds of standard algorithmic problems can be recast as learning tasks in specially-constructed MDPs \footnote{More precisely, to a family of tasks, specified by the instance size, we can associate a family of MDPs.}. We first give a trivial example of how an oracle problem can be transformed into an MDP. 

Consider Simon's problem, which is to identify the ``hidden shift'' $s$, given a function $f: \{0,1\}^{n} \rightarrow \{0,1\}^n$ that satisfies Simon's promise. We can imagine an agent that can input the queries $x$, and an environment that responds by outputting the value $f(x)$. The percept and action sets are thus bit-strings of length $n$. To encode the problem of finding the secret string, we can endow the agent with an additional set of actions of the form ``$guess-x$'' (for each $x\in\{0,1\}^{n}$), using which the agent can input a guess of the secret string, obtaining a reward only if the guess was correct (the returned percept can is not important and can \textit{e.g} be the guess that the agent had input). This indeed specifies an MDP. 

However, from a reinforcement learning perspective, this MDP is highly degenerate. Specifically, the rewards are immediate, the environmental transition matrices\footnote{Note that the transition function $T(s|a,s)$ can be understood as a collection of action-specific transition matrices $\{ P^a \}_a$.} which specify how actions influence state-to-state transitions are low rank, and it is fully deterministic. This degeneracy also means we can realize it using environmental maps which allow oraculization (see \ref{SI} for more details). Specifically, we can realize the unitary oracle  $\ket{x}\ket{y} \mapsto \ket{x} \ket{y \oplus f(x)}$, which allows a quantum  agent to obtain rewards exponentially faster -- and this is indeed the basic idea behind this work. 
Hence, this MDP is not particularly interesting from a RL perspective. 

In the following sections, we will show that Simon's problem can be transformed into MDPs which have more interesting structure, which is more typical of interactive RL problems.

\section{Interactive reinforcement learning problems}

\label{sec:main}

What deserves to be called a \textit{genuine} RL problem is a complex question, which we do not presume to resolve here.\footnote{In this work we treat the terms ``genuine'' and ``generic'' interchangeably.} 
Our objective is more modest, specifically to identify criteria which exclude certain MDP families -- those which \textit{directly} map to more specialized problems in quantum query complexity or computational learning theory. We refer to MDPs that satisfy our criteria as \textit{inherently interactive} RL environments. 

In particular, our desiderata for genuinely interactive MDP characteristics are thus:\vspace{0.2cm}\\
\indent$a)$ \textit{Rewards are delayed, and the MDP rewarding diameter (essentially, minimal number of moves between two rewarding events, \textup{e.g.} length of a chess game)
 should grow as a function of the instance size.}\\
\indent$b)$ \textit{At every step, the agent's actions should influence the states that are reached later, as well as the reward that the agent eventually receives.}\\
\indent$c)$ \textit{At every step, the optimal action should depend on the current state. (In particular, an agent that plays a fixed sequence of actions, without looking at the labels of the states, will be sub-optimal.)}\vspace{0.2cm}\\
\noindent The property $a)$ eliminates direct trivial phrasings of computational problems, and most direct translations of conventional COLT settings. 

The \textit{MDP rewarding diameter} refers to the maximal length of the shortest sequence of moves of the agent that lead to a reward, provided the sequence started from a state that is with bounded probability 
recurrently reached under every optimal policy.
This criterion demands that the (nearly) optimal agents must traverse long paths between rewarding events. 

To clarify this concept a bit, we can first consider deterministic environments. Here, the rewarding diameter is just the minimal number of moves between two rewarding events. If the environment is probabilistic, then, depending on the policy, the frequency of visiting each state may differ. We are only interested in states which are visited with a high probability under optimal policy (we may have bad agents which take wrong turns, or make unnecessary  loops, but this is not an inherent feature of the environment, and we do not care about events which are very rare).
In each such set of  ``not infrequent'' states (specific to each optimal policy), we can find the ``worst'' state, in the sense that it it is furthest away from a next rewarding move. 
The rewarding diameter is the shortest such a worst-case path length, minimized over all optimal policies. 

In other words, in an MDP with a rewarding diameter $d$, any optimal agent will, with constant probability, have to execute $d$ steps between two rewarding events \footnote{Note that simply demanding that the MDP has long paths does not suffice to eliminate otherwise pathological settings: one could simply add long paths to an otherwise immediate reward setting, which are never needed for optimal behavior. The rewarding diameter condition eliminates such possibilities.}.
Since we are interested in scaling statements, \textit{i.e.} statements regarding speed-ups with respect to (ever growing) families of MDPs, we additionally demand that the diameter grows as well.

Property $b)$ eliminates pathological MDPs  where only every $k^{th}$ move is (potentially) rewarding, and the actual moves the agent makes in-between do not matter. For example, this excludes constructions where one takes a small MDP and inserts meaningless ``filler'' transitions in order to artificially increase the rewarding diameter.

Finally, $c)$ eliminates deterministic maze settings, where optimal behavior only requires the repetition of a fixed sequence of actions. The problem of discovering the right sequence of actions is more closely related to computational learning theory, rather than reinforcement learning. 
The criteria as presented above are meant to convey certain intuitions about what properties ``genuinely interactive'' RL settings should have, considering both the agent and environment. These could nonetheless be, in principle, fully formalized in terms of the characterizations of the transition function and the reward function of the environment. For instance, the criterion b) captures a facet of an abstract property of the transition function of the MDP (which can be understood as a 3-tensor), prohibiting it to be of too low a rank.
Let us exemplify this on the case that b) is violated, meaning many actions lead to a same state-to-state transition. We can now view the transition function as a collection of current-state specific matrices, where each matrix specifies the subsequent state, given an action. 
Violation of b) directly implies that (at least one) of these initial-state-specific matrices can be low rank, as many actions lead to the same state. 

The criteria as listed are not fully independent, if viewed from the perspective of the characterization of the transition function of the environment, but are more meaningful from the characterization of optimal agents.
It is useful to point out one last, for this work important, consequence of property c). Property c) implies that the environment cannot be deterministic; indeed, in deterministic environments, an optimal agent need only store the sequence of moves it needs to perform, independently from the actual environmental state, and ``blindly'' repeat this sequence. Hence, the only way to ensure that an agent must develop an actual state-action specifying policy is to introduce random transition elements.

Since what is meant by ``genuine'' RL problem is subjective, and a matter of context, we do not proceed further with the full formalization of such (to an extent arbitrary) criteria. For instance, for our purposes it makes little sense to precisely specify implicit parameters of the criteria, \textit{e.g.} what is the slowest acceptable growth of the rewarding diameter, or, just how dependant (relative to some measure of correlations) to the subsequent states have to be to the chosen action of the agent.  Even though they are not absolutely rigorous, the properties $a) - c)$ can be clearly verified for the constructions that follow.
\vspace{0.2cm}\\

\section{Interactive RL environments that lead to exponential quantum speed-ups}

\label{exp:sec}

We will now construct MDPs that have two important properties: first, these MDPs satisfy the definition of an interactive RL environment given in Section \ref{sec:main}, and second, when these MDPs are realized by a quantum-accessible environment, it becomes possible for a quantum learning agent to achieve an exponential improvement over the best classical learning agent. 

In these examples, the quantum agent works by applying the oraculization techniques described in Section \ref{sec-oracul}, thus ``converting'' the task environment into a quantum oracle. 
This paradigm can be represented with the following picture:

\EQ{
\begin{split}
MDP &\xrightarrow{\textup{realized\ as}} 
\biggl( {\footnotesize {\textup{quantum-accessible} \atop \textup{environment $E_{acc}$}}} \biggr) \\
&\xrightarrow[\textup{to simulate}]{\textup{agent uses $E_{acc}$}} ( \text{quantum oracle } E_q ) \\ 
&\xrightarrow[\textup{queries to $E_q$}]{\textup{agent makes}} \biggl( {\footnotesize {\textup{quantum advantage for} \atop \textup{reinforcement learning}}} \biggr).
\label{eq:process}
\end{split}
}\vspace{0.cm}

For concreteness, we will work with an example based on Simon's problem \cite{1995_Simon}, which leads to an exponential quantum speedup. However, we mention that a similar construction is possible based on Recursive Fourier Sampling \cite{1997_Bernstein, 2008_Hallgren}, which has a recursive game-like structure that seems natural in the context of reinforcement learning, although it only leads to a superpolynomial quantum speedup. Furthermore, reductions to essentially any other oracle identification problems can be done analogously. 

In order to show a classical-quantum separation, one must separately prove two statements: that a quantum agent can learn in the given MDP efficiently; and that no classical agent can learn efficiently. The first property, an efficient \textit{quantum upper bound}, reduces to proving that quantum oraculization is possible for the given MDP. The second property is typically more challenging. In our examples, we will use a reduction technique, showing that even though the realized MDP allows more options for the agent, learning in the MDP is not easier than learning from the original Simon's oracle -- in which case, the classical-quantum separation is well-established.

\subsection{From oracles to interactive RL environments}
\label{constructions}
Consider standard oracle identification problems, where $\{ f_{s}: X \rightarrow Y \}$ is a specified family of boolean functions, where $s$ is the secret string $s\in \{ 0,1\}^l$ to be identified, and $X = \{ 0,1\}^m$  $X = \{ 0,1\}^n$.

For didactic purposes, we shall first provide an MDP construction $M_0$ which closely follows the underlying structure of the oracle identification problem. This directly translated MDP $M_0$ has none of the desired ``genuinely interactive'' properties described in Section \ref{sec:main}. However, we will provide one intermediary MDPs, $M_1$  which will satisfy properties $a)$ and $b)$. Finally, we will provide two more demanding modifications realizing  $M_2$  which satisfies the following \textit{global properties:}\vspace{0.2cm}\\
 $i)$ it satisfies the ``genuine interactive MDP''  desiderata $a), b)$ and $c)$. The MDP construction is thus done on an abstract level, without considering any of the specificities of the underlying oracle identification task.
 Following this, we will separately prove that for the case when $f_s$ are the functions satisfying the Simon's promise, the resulting MDP 
  $ii)$ maintains the classical hardness of learning, and\\ 
  $iii)$  reduces via the oraculization process to the standard Simon's problem oracle, leading to a quantum-enhanced learning efficiency.
  Overall this establishs a hard exponential improvement of quantum agents over any classical agent.
  
 \paragraph{Initial construction of $M_0$}

As mentioned, the structure of such a function can be embedded to a degenerate MDP, by choosing $X$ to be the action set, $Y$ to be the MDP state space (an action $x \in X$ then results in state $f_s(x) \in Y$), and by appropriately encoding $s$ into a reward function: one method to do so is either the expand the action space to include a complete set of ``guessing actions''. Alternatively, if $l=m$, then each query can also be considered a guess. Note, all choices made here will have impact on whether classical hardness of learning holds, as it provides differing options of the agent.  
This MDP satisfies none of the conditions $a)$-$c)$ from Section \ref{sec:main}. The optimal policy is thus constant (\textit{e.g.} the agent should always output the correct guess), and not particularly interesting.

\paragraph{Unwinding the MDP: constructing $M_1$}
\label{sec:unwinding2}
We can do better by using the fact that the elements of $X$ are $m-$bit strings, and we can interpret individual bits as actions. The action set is thus $\mathcal{A} =\{0,1\}$,
an MDP episode is a set of $m$ sequentially performed actions, corresponding to one query. This results in an MDP with a smaller action set, and of longer non-trivial paths, which are more natural for RL settings.
To maintain observability of the MDP, in general, each sub-sequence of a query should result in a unique environmental state\footnote{Note, in fully observable settings the state should contain all the necessary information the agent would at any stage need to be able to proceed optimally -- which, in general, means it should be able to recover the action sub-string input up to the given point. }, and one simple way to do so is to expand the state space to contain all action substrings of all lengths, so $\mathcal{S} = \bigcup_{j=1}^{m}  \{ 0,1\}^j \cup \{0,1 \}^n$. For simplicity, we will assume $l=m$, in which case, the only rewarding action sequence is the action sequence specified by the secret string $s$ (an example of a construction where $l \not=m$ is given in section \ref{SI}).
The resulting MDP we call $M_1$.
This simple modification ensures property $b)$, as the actions of the agent genuinely influence subsequent states, but more importantly ensures that the rewarding diameter is not constant, but $m$.

\paragraph{Augmenting the MDP with a stochastic action: constructing $M_2$}

The MDPs constructed thus far still do not satisfy the characteristic $c)$ we set out to fulfil. 
This characteristic of the MDP, amongst other consequences, requires a stochasticity of the transition function, on the relevant/rewarding part of the environmental space. 
The property $c)$ asserts that the optimal behavior must explicitly depend on the state label, and not just on the interaction step counter. Note that full determinism of an MDP necessarily violates $c)$. Hence, we need to modify the MDP such that the transition function becomes stochastic. Further, stochasticity must appear in the part of the space that must be visited by optimal agents\footnote{Without demanding this, we could easily introduce stochastic regions to an MDP which is never visited by the agent, which is a trivial, yet uninteresting solution.}. This can be achieved by a relatively simple trick:
the action space is augmented to contain the random-jump action $rg$, which lands the agent at a random state, somewhere in the first half of the query specified by the secret string $s$. Note, optimal behaviour now requires the agent to always choose the option $rg$ as this leads to the shortest time intervals without rewards on average, but also prevents the agent from ``blindly'' executing any sequence of action: the $rg$ jump lands in a random state in the rewarding path so the subsequent actions of the agent do depend on where the jump landed\footnote{The agent can in principle execute any action in any state, however the valid jump occurs only if the $rg$ move is executed at the ``zeroth'' level. To fully specify the MDP, we must specify what happens also when this action is executed in any other state. It will be convenient to define that such an action leads to an arbitrary sequence of states such that the normal query depth of $m$ is reached, which will ensure that the MDP is essentially strictly episodic, which simplifies the oraculization process.}.

\begin{figure*}
\centering
 \includegraphics[width=0.8\textwidth, trim=8.5cm 10.5cm 8.5cm 3.5cm,clip=true]{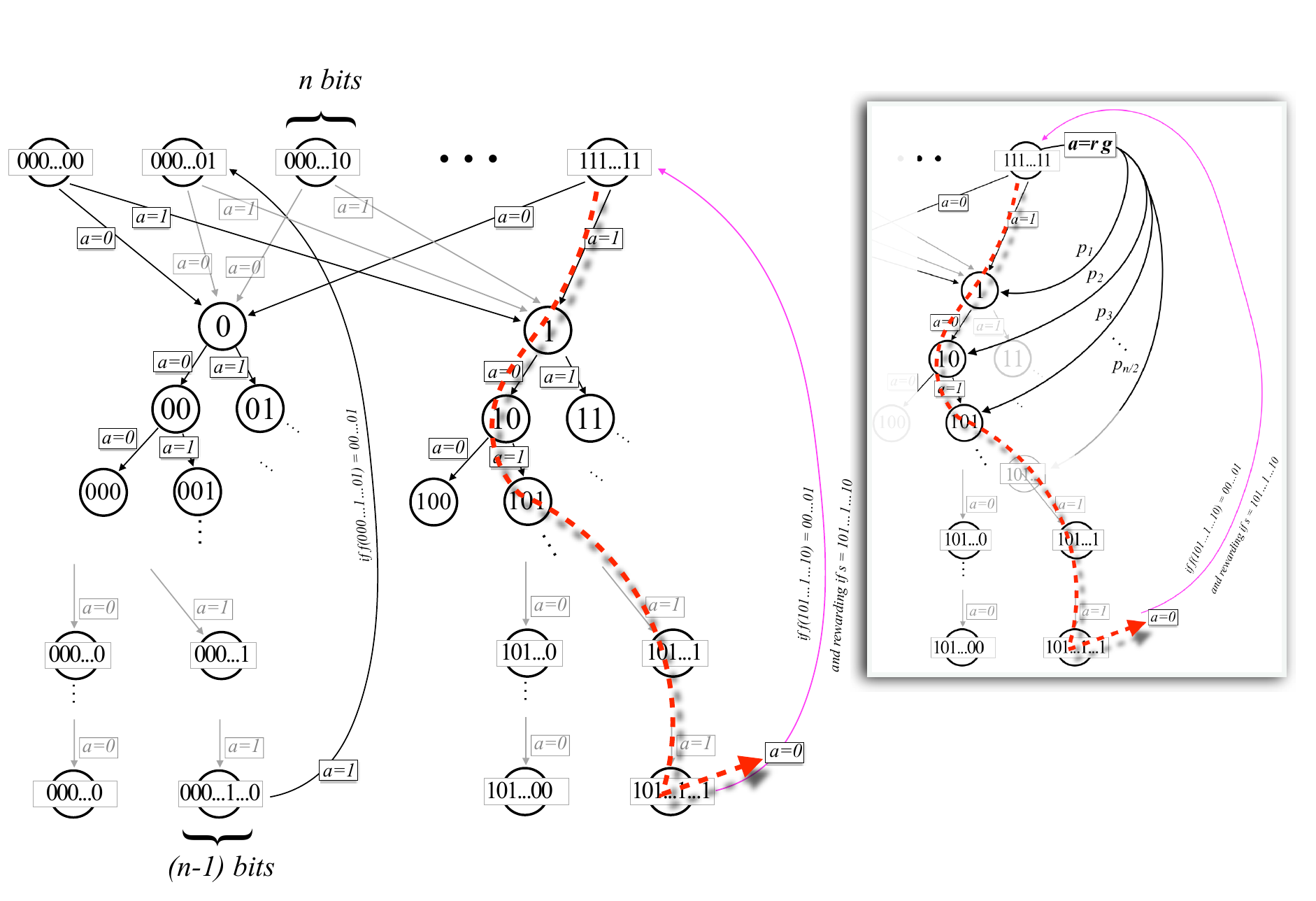}
\caption{\label{Simon} Illustration of $M_{1}$ and $M_2$.
In the deterministic case, the agent has 2 possible actions at each step, $\{0,1\}.$
 The actions form a tree of depth $n-1$, the last action causes a transition back to the zeroth layer. The path of $n$ moves encodes the input to the Simon's oracle, and the resulting state of the zeroth layer the output. 
 Each path is also interpreted as a ``guess'', and if the agent inputs $s$, the resulting transition is rewarded. The winning path, and rewarding transition are highlighted in red and pink. 
The inset figure presents the randomized version which allows the agent to land on a random element of the half-prefix of the winning path from any zero-layer state (state ``all-ones'' in illustration).
 }
\end{figure*}

An illustration of $M_1$ and $M_2$ for the example of the oracle for the Simon's problem is given in Fig. \ref{Simon}.

\subsection{Exponential speed-up from Simon's problem}
To prove that quantum agents can exponentially outperform any classical agent in (a random instance of) $M_2$ we will first prove that a classical agent requires an exponential number of interaction steps with environment specified by $M_2$ in order to get a reward even once (except with exponentially small probability).
\paragraph{Hardness for all classical agents}
For simplicity, we will work with a minor modification of Simon's problem, which we call the \textit{flagged} Simon's problem, where the query function also flags one bit, if the query is the secret shift, so: 
$f':\: \{0,1\}^{n} \times \{0,1\} \rightarrow  \{0,1\}^{n} \times \{0,1\},$ where 
\begin{equation}
f'(x,b) = (f(x), b \oplus \delta_{x, s}),
\end{equation}

where $f$ is some standard Simon's problem function, and $\delta_{x, s} = 1$ if $x=s$ and zero otherwise \footnote{In other words, the ancillary bit is flipped if the query is $s$.}.
Intuitively, it should be clear that learning $s$ given access to $f'$ is not (much) easier than having access to $f$: if $s$ is promised not to be ``all zeros'', then one can check whether some $s'$ is correct, also using $f$: simply evaluate $f$, for any $x,$ and $x\oplus s'$, and check if they are the same. For the full proof of hardness which also considers the case when $s$ is ``all zero'' we refer the reader to the Appendix, section \ref{flag:simon}.

Next, it is also relatively easy to see that if there exists any classical agent which efficiently learns in $M_1$ (the non-randomized version) for the Simons problem, then there exists an algorithm which solves the flagged Simon's problem as well -- the basic idea here is to simulate $M_1$ using nothing but a black-box access to the flagged Simon's oracle. The simulator simply returns the correct states given the actions (\textit{i.e.}, the complete sequence of actions input to this point), collect the actions until a query is complete, feed it into the oracle, and return the state and reward. 
Such a simulator combined with the learning agent is the algorithm which solves the oracular problem. 
This already proves that no classical agent can learn in $M_1$ efficiently.
Finally, we must take into account the randomized move option, which differentiates $M_1$ and $M_2$. To show hardness of $M_2$, we note that learning in $M_2$ is more difficult than learning in $M_1,$ where the agent is beforehand given the entire first half of the winning path $s$. In turn, this is as difficult as solving a $2n$ bit Simon's problem, where the first $n$ bits of $s$ are known.  It is relatively easy to see that solving this is not easier than solving a completely independent Simon's problem of size $n$ which is still exponentially hard. 
More precisely, these arguments can be used to prove the following result:
 \TH
 Any classical learning agent which can achieve $(poly(n)^{-1}, poly(n)^{-1})-$efficiency requires at least $O(2^{n/4}/poly(n))$ interaction steps in $M_{2},$ generated from the Simon's problem.
 \HT 
Full details of these proofs are given in the Appendix, section \ref{flagsim:leak}.

\paragraph{Efficiency for quantum agents}

The proof that there exist quantum agents which achieve optimal performance in $M_2$ can be provided in two steps.
First, one can show that there exist environmental realizations (\textit{i.e.} sequences of CPTP maps) of an environment specified by $M_1$, which allow oraculization, realizing one call to the standard Simon's unitary oracle  $\ket{x}\ket{y} \mapsto \ket{x} \ket{y \oplus f(x)}$, by using $O(m)$ interaction steps. This follows from the fact that the environment is essentially $m-$periodic, and from the fact that self-reversible realizations are always possible (see \cite{2015b_Dunjko} or section \ref{com:oracul} for more details).
Moreover, when the environment is fully observable, the oraculization can be done even without the environmental reversal, which yields a simpler oraculization process \footnote{This is elaborated in section \ref{simp:or} in more detail, but our results would hold also without these simplifications -- however, this also shows that there exist a more general set of quantum realizations of fully observable environments which allow oraculization, than what is possible in the case the environment is not fully observable.}.
Second, it is clear that $M_1$ can be understood as a sub-MDP of $M_2$, realized by blocking any agent from utilizing the randomized action $rg$. 
This also means that there exist CPTP realization of the environmental maps of the environment given by $M_2$ which match a quantum accessible realization of an environment given by $M_1$ on the subspace not containing the action subspace spanned by $\ket{rg}$ can be used by a quantum agent to realize the standard Simon's unitary oracle. In other words, a quantum agent can learn $s$ in an environment realizing $M_2$ by simply behaving as if it were in the environment given by $M_1$.
All in all, an agent with quantum capabilities can achieve perfect performance in $M_2$ using $O(m^2)$ steps (a multiplicative factor of $m$ comes from the fact that each oracular query corresponds to $O(m)$ interaction steps of the agent).
This proves the following main theorem:
  \TH
 Environments specified by MDP $M_2$, stemming from a function satisfying Simon's promise allow an exponential separation between classical and quantum $(\epsilon, \delta)-$efficient learning agents, as long as $\epsilon, \delta$ are not super-polynomially decaying. In particular, the separation holds for constant error and failure parameters. Finally $M_2$ satisfies all three criteria $a)-c)$ for MDPs with generic properties.
\HT

\subsection{Practical uses of quantum-enhanced RL}
The results presented so far prove that quantum agents can learn exponentially faster than their classical counterparts. 
While this has clear foundational relevance, it is also important to ascertain whether such results can be expected to influence reinforcement learning as applied in the real world.
One major concern is our use of oraculization, which requires the agent to interact with the environment in superposition: it is not clear whether this can be achieved in realistic settings. Another concern is the fact that our quantum speed-ups are obtained for very special environments, which may seem artificial and unrealistic. Here we briefly comment on these concerns. In particular, we argue that oraculization can be achieved in settings where an agent learns to play a game, by playing simulated games against itself (``self-play''). Furthermore, we argue that one can achieve superpolynomial quantum speed-ups on a somewhat more natural class of environments that resemble recursive games, based on the Recursive Fourier Sampling problem and its generalizations \cite{1997_Bernstein, 2008_Hallgren}.

\paragraph{The feasibility of oraculization}
First, in standard RL settings, the environments are classical, and macroscopic, which effectively prohibits useful oraculization.
However, many of the celebrated results involving RL deal with simulated, rather than real environments, and RL is used as a ``pre-training'' process.  One of the best examples is the AlphaGo system, in particular. the most powerful AlphaGo Zero variant \cite{2017_Silver,2017_Silver_b}, where the system is trained by utilizing simulated games: self-play -- essentially by playing one agent against a copy of itself -- before it was tested against human and non-human opponents.
Since such simulations are done internally, ``in the mind of the agent'', oraculization is clearly possible, as soon as sufficiently large quantum computers become available.
More generally, any RL setting which involves model-based learning \cite{2009_Russel}, where the learning agent constructs an internal representation of the external environment, presents a perfect setting for our results to be applicable. 

A second domain where our techniques may be applied is quantum RL in quantum laboratories: there the environment is manifestly quantum, and so techniques like register scavenging and register hijacking are possible, at least in principle. To elaborate on this, in recent years, there has been an increasing interest in utilizing machine learning techniques to mitigate various obstacles one encounters when complex quantum devices, such as quantum computers are built. Indeed ideas on how to use machine learning to help in achieving more efficient quantum fault tolerant computation, how to mitigate error sources, and more generally, ideas on how to use ML to build a scalable quantum computer similar have been put forward (see \textit{e.g.} \cite{2018_Dunjko} for a review). Some such ideas rely on reinforcement learning and it makes perfect sense to utilize, if possible, fully coherent methods, \textit{i.e.}, quantum-enhanced reinforcement learning \footnote{Naturally, for this to be feasible, at least a constant size quantum computer should be achievable, which is capable of running the quantum-enhanced algorithm -- it is intriguing to consider the possibility that such a process could be ``boot-strapped'' and made to correct itself, as an autonomous and intelligent and adapting quantum fault tolerant method.}.  

\paragraph{The kinds of MDPs that lead to quantum speedups}
Our second concern has to do with the still-rigid properties that the MDPs have to satisfy before quantum speed-ups can be obtained.
As a first response to this issue, we point out that, while the results we presented deal with Simon's problem, similar methods can be used for other problems as well. In the Appendix, section \ref{speed:RFS}, we show how the Recursive Fourier Sampling (RFS) problem can be used to provide super-polynomial separations \cite{1997_Bernstein, 2008_Hallgren}. 
In particular, we prove the following theorem:
 
   \TH (informal)
There exist families of MDPs, constructed on the basis of RFS problems, which allow a super-polynomial separation between classical and quantum $(\epsilon, \delta)-$efficient learning agents, as long as $\epsilon, \delta$ are not super-polynomially decaying. In particular, the separation holds for constant error and failure parameters.  These MDPs satisfy all three criteria $a)-c)$ for MDPs with genuinely interactive properties.
\HT

 RFS, in its original formulation \cite{1997_Bernstein}, assumes access to an $O(n\times \log(n))$-bit binary function $f$. The function $f$ satisfies rather complex nesting conditions, and the classical-quantum separation is in the identification of one bit, concealed in the specification of $f$. In this sense, the RFS problem does not fit in the paradigm of oracle identification tasks which yield hard RL problems, as, na\i vely, we are asked to distinguish between only two classes of functions. If only a correct guess is rewarded, as it would be the case in a simple lifting of RFS problems to MDPs, then already one attempt at guessing would already reveal the correct solution. However, in the formulation of RFS given in \cite{2008_Hallgren}, which studies generalizations of RFS, it is apparent that RFS can also be understood as the problem of identifying an $n-bit$ hidden string. The identification of this string can be achieved in \textit{poly}(n) given quantum access. In contrast, the classical bound for the identification of this string is super-polynomial. 
 
 Starting from this formulation, and by using constructions similar to those in section \ref{constructions}, we can recover environments, specified by MDPs where quantum access allows efficient learning. Further, we prove that the learning problem is still hard for classical learners. This turns out to be a bit more involved than for the case of MDPs based on Simon's problem, and is achieved using a lifting construction, which embeds smaller instances of RFS in larger instances. Using this, we show that the leaking of parts of the secret string still yields a problem harder than a fresh RFS problem of a smaller instance size. 
 The exact statements, proof and all constructions are extensively described in the Appendix, section \ref{speed:RFS}. 
 The constructions stemming from the RFS problem are particularly interesting because RFS exhibits certain self-similar features. Such features are reminiscent to features of learning we often encounter in real life, see the Appendix, section \ref{self:sim} for a discussion. 

Finally, while in this paper we have focused on \textit{provable} quantum speedups, it is worth taking a few moments to consider what kinds of problems might be good candidates for \textit{conjectures} of quantum speedups. 
Indeed, we can modify the MDPs described in this paper in various ways, such that our quantum algorithms can still be applied (with the same efficiency), and such that we might still plausibly conjecture that no classical agent can perform well (although we are no longer able to give a rigorous proof of classical hardness). 

One example of this has to do with the promise hidden in the underlying problem, which can be relaxed, thereby increasing the applicability of the underlying quantum algorithm for oracle identification. 

Another example has to do with embeddings of one MDP into another MDP. This is genuinely linked to the process of quantum oraculization, and is thus more interesting from our perspective.
In the process of oraculization, the agent can, for instance, ``ignore'' certain options, and recover a given oracle. 
 This is further discussed in the Appendix, Section \ref{upperB}, and one particular aspect is formalized in Lemma \ref{QuantumUB}. This states that whenever the restricting of an agent's actions leaves the agent operating in a sub-MDP which can be usefully oraculized, this opens the door for efficient quantum algorithms (although, a-priori nothing can be said about whether there also exist classical efficient algorithms). 
 
The above idea can be generalized further. 
Note that the restricting of the agent's actions (to realize useful sub-MDP) can be understood as a filter or interface, placed between the agent and environment, which, intuitively, rejects some of the agent's moves. 
But much more elaborate interfaces can be used, and such interfaces capture various notions of ``embedding'' of one MDP into another. This also expands the applicability of our results to all MDPs which embed any of the examples we have explicitly provided in this work. 
We leave a more extensive analysis of this options for future work.

\section{Discussion}
\label{sec:dis}
The presented constructions balance between three requirements which all have to be fulfilled to achieve the goals of this work: demonstration of better-than-polynomial speed ups for interactive RL tasks. First, it should be hard for a classical agent to learn in a given MDP, and moreover this should be rigorously provable. Second, the quantum agent should be able to usefully ``oracularize'' the provided environment under reasonable concessions. Third, the MDP should be \textit{interesting}, that is, have properties which are quintessential to RL. 

The second and third requirement are in fact in strong collision: interesting RL settings involve long memories, and dependencies which vary in length, all of which interfere with the agent's efforts to ``oracularize'' the environment. 
To resolve this collision, in this work we settled for what is arguably the simplest possible solution: we constructed MDPs with randomness that occurs only along the rewarding path.
This has a few consequences, e.g. the optimal strategy (which uses the stochastic part of the MDP) is not much better than the strategy which resides in the deterministic part of the MDP: $(n\times l)$ vs $(n\times l - n/4)$ steps between rewards. 

In order to increase this separation, the agent would have to classically operate in the randomized section of the environment for longer, in which case effectively quantizing just the deterministic part of the environment would lead to a less of advantage. Alternatively, one could attempt to  genuinely quantize/oraculize also the random parts of the environment, however this leads to stochastic oracles whose utility is still not fully understood. {A few results in this direction suggest that quantum improvements in such scenarios may be difficult, as in many cases noisy or randomized oracles offer no advantage over classical oracles \cite{2008_Regev,2014_Harrow}.  }

{As a possible route of future research, one may attempt to consider MDPs with a larger stochastic component, by considering settings which do not correspond to standard MDPs.} For instance, if the environment is allowed to be time-dependent, then one {could} consider the task consisting of two phases -- a deterministic phase, where quantum access is used to learn useful information, \textit{a key}; and a stochastic phase, where the key is necessary to successfully navigate the environment. This would entail a full formalization of ideas of hierarchical learning and information transfer, also discussed briefly in section \ref{self:sim}. As an example of such learning, one {could} consider the notions of information transfer from one environment to another, where already constant separations between learning efficiency may lead to settings where the agent behaves optimally in the limit, or no better than a random agent which learns nothing. 

To exemplify this, consider a \textit{nested mazes} environment: a sequence of ever larger mazes $E_0, \ldots, E_k$ where the $E_l$ consists of $E_{l-1}$ glued to a new maze (the exit of $E_{l-1}$ is the entrance to the new maze), called an appended maze, denoted $E'_{l}$. In each maze $E_l$ only the final exit is rewarded. Because of this, learning $E_l$ is not equal to $l$ independent instances of learning appended mazes, but is significantly harder.
We assume the appended mazes are roughly of the same size (and take the same time to traverse), and $E_0 = E'_1$.
Now we can define a \textit{growing maze} setting, where an agent is kept in $E_{l}$ for some number of time steps $\tau_l$, before it is moved to $E_{l+1}$. In such a scenario, even a constant difference in learning speed can become magnified exponentially in $l$.
An agent which can manage to learn each appended maze $E'_l$ in time $\tau$ can avoid ever having to earn a maze of increased size: it learns $E_1,$ and solves $E_2$ by applying first the solution of $E_1$, which brings it to the beginning of the appended maze, which is of constant size. 
Later, the agent has the simple recursive step: to solve $E_l$, it executes the solution of $E_{l-1},$ which brings it to the new, but constant sized instance. 
Assuming that each maze can be traversed in say $\kappa$ steps, as long as $\tau_{l} \geq \kappa\times \tau$, the agent will be successful each time, effectively never having to tackle a larger maze.
This is a simple example of transfer learning, where knowledge in one domain is utilized in the next.
In contrast, any agent which requires more than roughly $\tau_l / \kappa$ steps to learn the appended mazes, will have to learn the large mazes from scratch. This will imply exponentially worse success probabilities in $l$, rapidly converging to the performance of a random agent.  

Similar effects {could} be achieved in partially observable MDP cases, however, there the optimal policies may not be constant, but rather depend on the entire history of interaction. 
{Finally, it would be particularly interesting to identify the possibilities of speed-ups in RL settings which do not utilize a reduction onto oracle identification problems, but deal directly with environmental maps.}

\noindent\textbf{Acknowledgements}
VD and JMT are indebted to Hans J. Briegel for numerous discussions which have improved and influenced many parts of this work.
The authors wish to thank Shelby Kimmel for initial discussions, and Stephen Jordan, Scott Glancy and Scott Aaronson for helpful feedback.
VD also thanks JMT and Hans J. Briegel for their hospitality during his stays.
VD acknowledges the support from the Alexander von Humboldt Foundation. 
JMT acknowledges the support of the National Science Foundation under Grant No. NSF PHY11-25915. 
Contributions by NIST, an agency of the US government, are not subject to US copyright.
The authors acknowledge funding from ARL CDQI.
This material is based upon work supported by the U.S. Department of Energy, Office
of Science, Advanced Scientific Computing Research Quantum Algorithms Teams program.

\newpage 

\bibliographystyle{plain}

\begin{widetext}
\pagebreak
\newpage
\end{widetext}

\appendix

\section{Oraculization of quantum-accessible environments}
\label{SI}
\subsection{Constructing the oracle $E_q$}
\label{com:oracul}

Consider an agent facing a quantum accessible (deterministic, $\eta-$episodic) environment, where the overall setting additionally allows the agent to intermittently interfere with the ancillary workspace via processes called \textit{register scavenging }and \textit{register hijacking} \cite{2015b_Dunjko}.

In Section \ref{sec-oracul}, we claimed that the agent can utilize approximately $5 \eta $ interaction steps
 with the environment in order to simulate a particular type of oracle:
\EQ{
E_q: \ket{a_1, \ldots, a_\eta}\ket{y} \mapsto \ket{a_1, \ldots, a_\eta}\ket{y \oplus R(a_1, \ldots, a_\eta)}  \nonumber,}
where $R(a_1, \ldots, a_\eta)$ is the reward value obtained by the agent once the agent executes the sequence of actions $a_1, \ldots, a_\eta$,\footnote{Note that this is a well-defined quantity only in deterministic environments, where the action sequence deterministically specifies the corresponding state sequence, and reward values.} and $\oplus$ denotes addition in the appropriate group.

We now explain, at an intuitive level, how this can be accomplished. Notice that if the agent simply performs the sequence of actions $a_1,\ldots,a_\eta$, this results in a state of the form $\ket{a_1,\ldots,a_\eta} \ket{s_1,\ldots,s_\eta} \ket{R} \ket{j_1,\ldots,j_\eta}$, where $s_1,\ldots,s_\eta$ are the percepts returned by the environment, $R$ is the resulting reward, and $j_1,\ldots,j_\eta$ represent the contents of any auxiliary quantum subsystems that are retained within the environment. In order to simulate the oracle $E_q$, the agent needs to gain control of the auxiliary states $\ket{s_1,\ldots,s_\eta}$ and $\ket{j_1,\ldots,j_\eta}$, and then erase or ``uncompute'' them. 

For this to be possible, the agent needs extra access to the environmental registers, and uncomputing must be feasible. 
The access is ensured by assuming scavenging and hijacking options, which were defined specifically for this purpose.
The uncomputing of the environmental registers carries a different problem. In general, it would seem to entail a need for an access to a \textit{reversed} environment, which implements the Hermitian adjoint of whatever unitary map is overall realized by the environment.
However, this assumption is not as problematic as it may seem: 
the implementation of any classically specified environment can be realized by a mapping where the ancillary state $\ket{j_1,\ldots,j_\eta}$ is equal to the percept specifying state $\ket{s_1,\ldots,s_\eta}$, since at each step the sequence of previous states/percepts and actions fully specifies the subsequent percept/state (ignoring probabilistic environments for the moment).
Further, at each step, barring the final rewarding step, the environment must produce the subsequent percept, given the current history of percept/action transitions. This can be realized as a controlled-unitary, which, conditioned on the states of percept/action containing registers, rotates a fresh ancillary action register to an appropriate action. Note that each such controlled unitary acts on separate target registers. Each such controlled unitary can thus be represented by a block-diagonal operator of the form $U_l =\sum_{h} \dm{h} \otimes U(h),$ where $h$ specifies a history, and $U(h)$ rotates a fiducial state $\ket{0}$ to the appropriate action state (determined by $h$). Each $U(h)$ thus needs to act non-trivially only on a two-dimensional subsystem.Consequently, $U(h)$ can be chosen such that it is Hermitian, or rather, self-inverse: $U(h)U(h) = \mathbbmss{1},$ which renders the entire operator $U_l$ self-inverse. 
Further, since each $U_l$ has differing target registers (but overlapping control registers), all operators of the environment (barring the rewarding operation) can commute. 
This means that the environmental transition map can always be implemented in a self-reversible fashion, which will allow uncomputation by simply running the environment twice.
In summary, the overall process is described as follows.
 The agent utilizes the first $\eta$ steps to input some action sequence of length $\eta$ (collecting $\eta$ percept states), while collecting the memory the environment traces out, by using \textit{scavenging}. Scavenging also costs $\eta$ steps, as $\eta$ subsystems are collected.

Following this, using hijacking ($\eta$ steps) and the fact the environment is self-reversible, the agent uses $\eta$ steps to ``un-compute'' the percept responses of the environment. Importantly, this can be done in a manner which does not un-compute the reward value.

Thus running the same interaction sequence twice can be used to ``uncompute'' unwanted information, whereas the information we wish to keep we need to protect using so-called hijacking.

 The last scavenging round ($\eta$ steps) collects the actions. 
$E_q$ is referred to as \textit{the oracular instantiation of the environment}, and each invocation of this oracle is counted as $5 \eta $ interaction steps. 
See \cite{2015b_Dunjko} for details on the construction.

The access to such an effective oracle $E_q$ was used to obtain a quadratic quantum advantage for quantum learning agents in \cite{2015b_Dunjko,2016_Dunjko}.
Note, in the case the task environments are constructed (e.g. in model-based learning settings \cite{2009_Russel}, where the agent internally constructs a simulation of the environment), the internal construction process can directly realize the oracular instantiation.  An environmental setting where the agent can choose to interact with the environment $E_c$ (any from the set $\{E_i \}$  where each $E_i$ is a sequence of CPTP maps, realizing the same input-output specification of the task environment under classical access) or $E_q$ (either via simulation or construction), we call \textit{a controllable environment}.

\subsection{Simple techniques for reversing the environment}
\label{simp:or}

A critical step of oraculization is the erasing of the environmental responses $(s_1, \ldots, s_k)$. In general, the states returned by the environment depend on the previous actions of the agent, and this implies that the corresponding registers get entangled under quantum access. To obtain the desired oracle $E_q$, this state information should be purged. In general, this requires the ``uncomputation'' of the state information, which requires reversing the dynamics of the environment (which is possible if the environment is implemented in a self-reversible fashion,  as discussed in the previous section, which we assume here\footnote{Note, this is an assumption on the implementation of the environment, not its specification. All classically specified environments admit a self-inverse implementation.}).

For certain types of environments, simpler techniques can be used to reverse the dynamics of the environment, which does not require running the environment twice to un-compute the environmental responses.

 For example, in the environments that we construct in this paper, the state information consists of the sequence of actions performed by the agent: if the agent performed the sequence of actions $0,1,1$, it is in the environmental state $\ket{011}$ (see Fig. \ref{MDP2} ).
 (Note, the state information is required to render the task environment fully observable, and for this, it would suffice to have all the state labels unique. The choice of the state labels collecting explicitly the path the agent took is but one possible choice.
 However, it is a particularly convenient choice.) 
 
 Consider now the overall state realized by the agent-environment interaction:
 \EQ{
 \ket{\psi} = \ket{a_1,\ldots,a_\eta} \ket{s_1,\ldots,s_\eta}_{1} \ket{R} \ket{s_1,\ldots,s_\eta}_{2}
 }
 After a scavenging step, this entire state is held by the agent.
 
In the case that each state is exactly equal to the sequence of performed actions leading to it, so 
$s_{k} = (a_1,a_2,\ldots,a_k)$, the deleting of registers $1$ and $2$ above can be done by the agent itself, with no need re-run the interaction with the environment.

\section{Exponential speed-up from Simon's problem}

\subsection{The flagged Simon's problem}
\label{flag:simon}
In the process of embedding an oracle identification problem into an MDP, one must decide how to encode the correct guess of an oracle into a reward. One way is to separate ``query-actions'' from ``guess-actions'', as is illustrated in the constructions stemming form the Recursive Sampling Problem we give later in this Appendix. 
However, when the number of oracles matches the number of possible queries, it is more interesting and natural to encode the query and guess in the same structure, by rewarding the query input which corresponds to $s$. Such an MDP encodes a slightly modified black-box function, which outputs a flag, if the query is actually equal to $s$. More formally, it encodes the function $f':\: \{0,1\}^{n} \times \{0,1\} \rightarrow  \{0,1\}^{n} \times \{0,1\},$ where 
\begin{equation}
f'(x,b) = (f(x), b \oplus \delta_{x, s}),
\end{equation}
where $f$ is the standard black-box function.\footnote{In other words, the ancillary bit is flipped if the query is $s$.} 
When the underlying problem is the Simon's problem, we call this modification the \textit{flagged Simon's problem}.
 
We prove that the flagged Simon's problem is not (significantly) easier than the original form.

 \LE \label{FSP}
 Flagged Simon's problem has an exponential classical lower bound.
 \EL
 \begin{proof}

 We again prove this via simulation. Note that if we assume that $s \not= 0\ldots0$, given access to Simon's oracle, we can easily check whether some given string $x$ matches the secret string $s$.  One simply queries Simon's oracle on two points $t$ and $t\oplus x$, and checks whether the output is the same.
 Suppose now that we have an algorithm $A$ which finds $s$ for the flagged Simon's problem, under the promise that $s\not=0\ldots0$, using $T$ queries.
 Now, given an oracle for the (original) Simon's problem, we use the algorithm $A$, and introduce a simulator, which for each query of $A$ outputs the value $f(x)$ (by direct query to the Simon's oracle), and performs a check if $x=s$ as described earlier. 
 After $T$ queries given by $A$ (and $2 T$ queries actually performed by the simulator), if $s \not  = 0\ldots0,$ then one of the checks confirmed a query by assumption of correctness of $A$. If not, we output the guess that $s=0\ldots0$. This proves that there cannot be an algorithm which learns the secret $s$ of the flagged Simon's problem in $ T \in O(2^{n/2-1})$ steps with zero-error. 
 
 For the case of randomized algorithms which can err, assume $A$ outputs the correct string with polynomially bounded failure probability $p$.
 The process we described for the deterministic error-free case is repeated $c$ times. If all fail, we output $s=0\ldots0$.
 The probability of this being incorrect is $p^c$. Since $1-p$ is at most polynomially decaying in $n$, overall we have that flagged Simon's problem does not allow a randomized algorithm which identifies $s$ with polynomially bounded error probability, using fewer queries than $\Omega(2^{n/2-1}/poly(n)).$
  \end{proof}

\subsection{Classical lower bound for the randomized  MDP $M_2$ stemming from Simon's problem}
\label{flagsim:leak}
As explained in the main text, need to prove that solving the (flagged) Simon's problem given access to the fist half of the secret $s$ cannot be (radically) easier than the problem without this information leak.

We prove this by embedding an $n-$size instance into a $2n$ instance, followed by a uniformization procedure, which ensures average case hardness. 
We begin with a technical claim regarding the relationship of two functions satisfying Simon's promise with the same shift.
\LE\label{unif2}
Let $f$ and $g$ be two $n-bit$ functions, both satisfying Simon's promise with the same $n-$bit string $s$.
 Then there exists a permutation $h \in \{0,1 \}^{n} \rightarrow \{0,1 \}^n$ such that $f(x) = h(g(x))$.
Conversely, if there exists a permutation $h$ such that $f(x) = h(g(x))$ and $g$ is satisfying Simon's promise with the $n-$bit string $s$, then so is $f$.
\EL
\begin{proof}
If either of the functions $f$ or $g$ are permutations, then both directions are trivial (e.g. $h(x) = f(g^{-1}(x))$).

Suppose $f$ and $g$ satisfy Simon's problem with the same shift $s$.
Then $f$ and $g$ act as constants on the same pairs of input bit-strings, as
\EQ{
f(x) = f(y) \Leftrightarrow x=s\oplus y \Leftrightarrow g(x) = g(y).
}
They can only differ in what values are attained for each pair, that is, by a permutation on the equivalency class $x \tilde y  \Leftrightarrow x=s\oplus y.$ This is the permutation $h$.

To prove the converse, note that since $h$ is a permutation, so injective and invertible, we have that $f(x) = f(y)   \Leftrightarrow h^{-1}(f(x)) = h^{-1}(f(y))$. But then
\EQ{
f(x) = f(y)  \Leftrightarrow h^{-1}(f(x)) = h^{-1}(f(y))  \nonumber \\ \Leftrightarrow h^{-1}(h(g(x))) = h^{-1}(h(g(x)))  \nonumber  \\  \Leftrightarrow g(x) = g(y)  \Leftrightarrow x=s\oplus y.
}

\end{proof}

Now we can prove the hardness of Simon's problem with information leak.
\LE
The Simon's problem of size $2n$, where the oracle additionally leaks the first half of the string $s$ has classical query complexity lower bound of $\Omega(2^{n/2})$.
\EL
\begin{proof}
Let $f_1$ be the function corresponding to an $n-$sized instance of the Simon's problem, with secret string $s_1$.
To raise this to an $2n$- sized instance, we choose another $n-$sized instance with the (known) function $f_0$ and string $s_0$, and define the concatenated function $f(x_0 \circ x_1)  = f_0(x_0) \circ  f_1(x_1) $. $f$ is clearly a $2n-$sized instance of the Simon's problem, with $s = s_0 \circ s_1$. 
To uniformize the construction, by Lemma \ref{unif2} it will suffice to compose $f$ with a uniformly chosen permutation of the size $2n$, which generates a uniformly chosen $2n$- sized instance with the secret string $s$.
\end{proof}
  By combining the hardness of Simon's problem with leak with the reduction in the proof of Lemma \ref{FSP} showing the hardness of the flagged variant, we analogously get the hardness of the flagged version under leak.   
  
 \LE \label{FSP2}
 Flagged Simon's problem has an exponential classical lower bound, even if half of the secret shift is provided.
 \EL

\section{Super-polynomial speed-ups from Recursive Fourier Sampling}
\label{speed:RFS}
\subsection{Recursive Fourier Sampling}
Recursive Fourier Sampling (RFS) \cite{1997_Bernstein} is an oracular problem in which super-polynomial separations between classical and quantum algorithms can be obtained.
{The problem owes its name to its recursive structure, which builds on the basic, unit depth instance.}
We assume a computable, binary function $f: X \times X \rightarrow \{ 0,1\}$ (inner product in the original Bernstein-Vazirani variant of RFS, generalized to many others in \cite{2008_Hallgren}).
The depth-1 instances are specified as follows: given access to an oracle evaluating $\mathcal{O}(x) = f(s,x),$ identify $s$, where $s$ is a \textit{hidden secret string}.
To achieve deeper structures, the problem is composed.
This leads to an $2^n$-arry symmetric tree construction, of depth $l$. 
In what follows, we adhere to the formalization as given in \cite{2008_Hallgren}, and we choose to work with bit-strings, for concreteness.

To each vertex of the $v$ of the $2^n$-arry tree we assign the local label $x_v \in X = \{0,1 \}^n,$ where the root has label $\emptyset$ (and is at level 0). Further, to each vertex we assign the path-label specified with, for a vertex at depth $k$, the sequence $\mathbf{x}_v = (x_{pa^{k}(v)}, \ldots x_{v}) \in X_{\Sigma} = \bigcup_{k=0}^{l-1} X^k,$ (where $pa(v)$ is the label of the parent of the vertex $v$, $X^k$ is the $k^{th}$ Cartesian power of $X$  with $X^{0} = \emptyset$, and $pa^k(v) = pa\circ pa \cdots pa(v)$ is the $k^{th}$ ancestor of $v$.) designating the unique path from the root to the vertex in question, using local labels of each vertex along the path. The label $ \emptyset$ is both the path-label and the local label of the root.

To each vertex we also assign a hidden string of length $n$, specified by a secret-string function defined on the path-labels: $s: X_\Sigma \rightarrow X$.

We are given access to the oracle for the generalized RFS (gRFS) problem\footnote{As noted, in the original RFS problem the function $f$ is the inner product, and the separation is based on the Hadamard transform. This has been since greatly generalized in \cite{2008_Hallgren}.} $\mathcal{O}_{gRFS}: X_{\Sigma}  \rightarrow \{0,1, \bot \},$ defined with:
\EQ{
\begin{split}
&\mathcal{O}_{gRFS}(x_1, \ldots, x_k, a) =\\
& \begin{cases} 
f(s(x_1, \ldots, x_{k-1}),x_k),\,
 &\! \! \! \! \textup{if } a=s(x_1, \ldots, x_k) \label{no1}\\
\bot,\,
 & \! \! \! \! \textup{if } a\not=s(x_1, \ldots, x_k), \end{cases}\\
  \end{split}}
  \textup{if} $0<k<l$, \textup{and}
  \EQ{
\mathcal{O}_{gRFS}(x_1, \ldots, x_l) &= f(s(x_1, \ldots, x_{l-1}),x_l)\  \textup{otherwise.}\hspace{6.85cm} \label{no2},
}
In other words, for the leaves of the tree, we can (indirectly) access the secret strings of the leaves' parents.
{Since we have chosen to work with bit-strings specifying both labels and secrets, Eqs. (\ref{no1}) and (\ref{no2}) contain an ambiguity as identical inputs can be interpreted as instances of a leaf-query $O_{gRFS}(x_1,\ldots,x_l)$, or as instances of a penultimate layer query $O_{gRFS}(x_1,\ldots,x_{l-1},a)$.
There are a few options on how to resolve this technicality, and in the subsequent constructions, we shall use one additional bit specifying whether we are requesting a leaf or a parent-of-leaf query. }

To access the hidden values of any vertex whose children are not leaves, we, generally, need the secret strings of $O(n)$ of the children first.\footnote{This is easy to see when $f$ is the inner product, as choosing all the children with labels corresponding to the canonical vectors returns exactly the secret string of the parent, bit-by-bit. Also, this is the reason why the scaling is $n^{log(n)}$ for the classical algorithm -- there are $n$ canonical vectors, and the tree is depth $log(n)$.} 
The oracle (or the root, if you will), hides one bit $b_{\emptyset}$, which is revealed given the secret string of the root:\vspace{-.2cm}
\EQ{
\textup{if } k=0, \mathcal{O}_{gRFS}(a) = \left\lbrace {\bot\ \textup{if }a\not= s(\emptyset)\atop b_{\emptyset},
 \textup{if }a= s(\emptyset) }\right.
}\vspace{-.2cm}\\
Intuitively, to access the hidden value $b_\emptyset,$ we need the secret string of the root,
which in turn requires $O(n)$ secret strings of its children, and each of those needs the same, and so on recursively. 
The standard gRFS problem is the computation of the single bit $b =\mathcal{O}_{gRFS}(s(\emptyset))$, with bounded error probability. The tree depth $l$ is chosen to be $O(\log(n))$ to realize instances with the superpolynomial separation.
While the results of  \cite{2008_Hallgren}  focus on returning the single bit value, the quantum algorithm employed actually returns the entire bit-string
$s(\emptyset)$ utilizing $O(n^{2 \textup{log}((8/\delta)\textup{ln}(8/\delta)) })$  queries, where $\delta$ is a constant\footnote{This constant depends on the function $f$ (and dispersing unitary $U$ which can be used to solve the corresponding problem), but for simplicity this can be the inner product, so in $n^{O(1)}$ queries.}.
{To emphasize the fact we consider the problem of returning the entire secret, we refer to the constructions above the \textit{recursive hidden secret problem}.}

The (classical) lower bound of the query complexity is established for the problem of identifying the final bit $b$. However, the problem of finding the entire sequence $s(\emptyset)$ is harder: if an algorithm using $n^{o(log(n))}$ (less than $n^{log(n)}$) can only guess with probability bounded away by $n^{-\Omega(log(n))}$, then no algorithm with running time  $n^{o(log(n))}$  can output the bit sequence $s(\emptyset)$ with probability above $n^{-\Omega(log(n))}$ (as it would cause a contradiction).
The quantum oracles for this problem are standard ``bit-flip'' oracles. 
{In the specification of the classical oracle $\mathcal{O}_{gRFS}$ the input size may vary, which is not standard in the case of quantum oracles. This is resolved by either using access a family of oracles of varying input sizes which can be called, or, alternatively we can introduce an ancillary symbol to the input space, to ``void'' parts of the input in the case of smaller input sizes.  }

\label{sec:main2}
\subsection{Super-polynomial separation from Recursive Fourier Sampling}

As explained in the main text, the overall idea is to construct MDPs, which can be realized by task environments, which the agent can (via oraculization) ``convert'' into useful quantum oracles. 

Specifically, we construct environments that lead to RFS-type oracles. Moreover, we prove that no classical agent can learn efficiently in the given environments, as this would lead to a contradiction with the optimal performance of classical RFS solving. 

For didactic purposes, we shall first provide an MDP construction $M_0$ which closely follows the underlying structure of the RFS problem. This directly translated MDP $M_0$ has none of the desired ``generic'' properties described in Section \ref{sec:main}. However, we will provide two intermediary MDPs, $M_1$ and $M_2,$ which will satisfy properties $a)$ and $b)$. Finally, we will provide two more demanding modifications realizing  $M_3$  which satisfies the following \textit{global properties:}\vspace{0.2cm}\\
 $i)$ it satisfies the ``generic MDP''  desiderata $a), b)$ and $c)$, \\
  $ii)$ it maintains the classical hardness of learning, and\\ 
  $iii)$ it reduces via the oraculization process to the same RFS-type oracle, leading to a quantum-enhanced learning efficiency.\vspace{0.2cm}\\

\subsection{The basic construction $M_0$}

Given an RFS oracle of tree-depth $l$, over $n-$bit labels, the simplest MDP $M_0$ would, as the action set have all the possible queries to the oracle, and the possible outputs of the oracle as the set of the states of the environment (MDP).
The only rewarding action is the action with the label $(s(\emptyset))$ executed given any state. This MDP satisfies none of the conditions $a)$-$c)$ from Section \ref{sec:main}. The optimal policy is thus constant, and not particularly interesting.

\subsection{Unwinding the MDP}
\label{sec:unwinding}

We can do better by restricting the action set to the set of $n$-bit strings (recall queries to the RFS oracle can comprise path-labels, so sequences of $n$-bit strings), with an additional action denoted ``$q$'' (for ``query'') used to denote that the agent wishes to execute a query -- this controls the size of the input in the RFS oracle setting. Intuitively, the agent can now ``input'' the desired query, one $n$-bit block at a time, marking the end of the query with $q$.
Note that, in this case, to maintain full observability of this MDP $M_1$, the state space includes path-labels, telling the agent which moves it did thus far. 
This MDP $M_1$ has non-trivial paths, but the optimal policy is still not genuinely delayed as the two action of $ (s(\emptyset)),q) $ executed at the root lead to the reward and reset the problem -- i.e. the rewarding diameter is constant (two).

\begin{figure}[H]
\centering
 \includegraphics[width=0.5\textwidth, trim=6.2cm 6cm 15cm 6cm,clip=true]{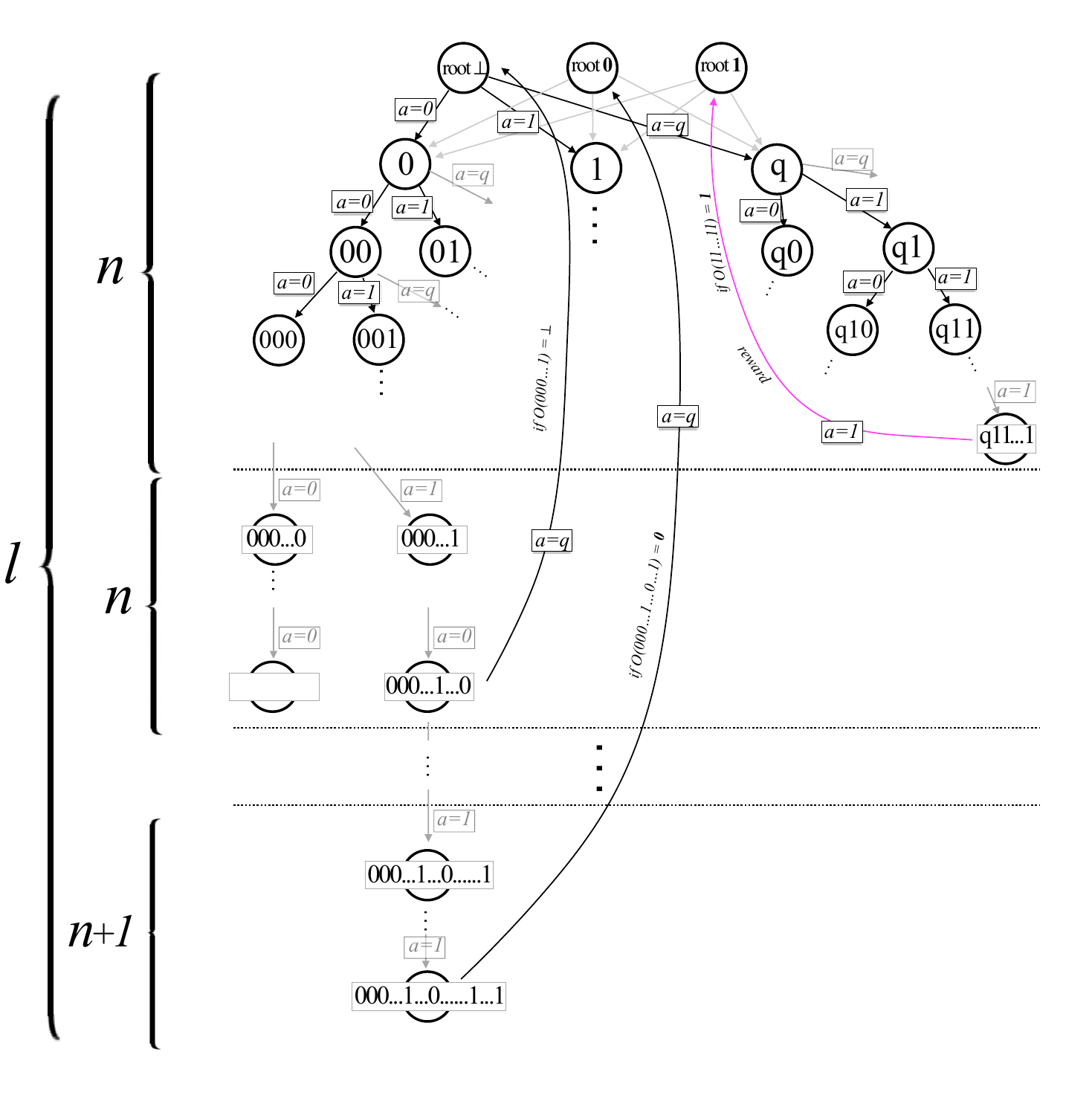}
\caption{\label{MDP2} Illustration of the MDP $M_2$.
The agent has 3 possible actions at each step, $\{0,1,g \}.$
 The tree has $l,$ $n-$deep sections. Each  $n-$deep section effectively loads the next $n$-bit string. The $q$ action is used to denote that the last sequence of $n$ actions ends the query, effectively specifying the input length. The labels of the directed edges encode the criterion under which the transition occurs: e.g. a move to ``root 1'' occurs from some leaf if that particular query would result in the output $1$ from the oracle. Performing the ``query'' action at the root level signifies that the agent will attempt to produce a guess of $s$. 
The maximal input length is $n\times l+1$. 
{The last layer accepts $n+1$ actions, as the last action resolves the ambiguity between oracular queries to the leaves of the RFS problem, and queries to the penultimate layer (with the appropriate guess). }
 The only rewarding transition is given in pink.  Invalid actions (not shown) lead to irrelevant sub-trees of maximal depth, linking back to any root state. }
\end{figure}

The problem of constant rewarding diameters can be circumvented by further restricting the action set to just $\mathcal{A}=\{0,1,q\}$, and instead of inputting labels, the agent inputs bit by bit. 
The states are the thus-far input actions of the agent, to ensure full observability. Non-legal inputs (e.g. $q$ executed where the number of input bits is not a multiple of $n$) lead to subtrees where all paths are of the same length (e.g. $l\times n$), which revert to the $\bot$ state of the root, without reward. Such an MDP is denoted $M_2$ and it is satisfying both the requirements of $a)$ and $b)$, as the rewarding action sequence is of length $n+1$. A (partial) illustration of $M_2$ is given in Fig. \ref{MDP2}.
{To explicitly separate oracle queries from ``guesses'' of the secret string $s$, if the agent performs the $q$ action at the root level, this encodes a guess, and only a correct guess yields a reward. This follows the construction we provided for the Deutsch-Jozsa problem example, and in the section \ref{exp:sec}, exemplified on a different oracle problem, we will show an alternative way of encoding guesses.}

{As mentioned earlier, the standard oracular formulation of the RFS setting (Eqs. (\ref{no1}) and (\ref{no2})) have an ambiguity when the secrets and labels come from the same set. Specifically, the queries at the leaves of the tree $O_{gRFS}(x_1,\ldots,x_l)$, and the queries at the penultimate layer of the tree (parents of leaves) $O_{gRFS}(x_1,\ldots,x_{l-1},a)$, both have the same form. To resolve this (purely technical) ambiguity, the last layer accepts $n+1$ actions, and the very last bit is used to designate a leaf or a parent-of-leaf query.}

More generally, given any MDP, with a large action set, where the action set can be encoded as a sequence of a fewer number of  actions, and where the state space is enlarged as to respond to the intermediary actions by the elapsed action sequence (this is necessary to maintain full observability of the environment), we can always increase the MDP's rewarding diameter, by reducing the action space. 
Observe also that the oraculization procedure constructing/simulating $E_q$ from multiple calls to $E_{acc},$ effectively achieves a reversed process. In this case, the action sequences are promoted to individual actions, which combinatorially increases the action set, the intermediary states are discarded, and, ultimately, and a constant MDP reward radius is achieved. 
Both processes we refer to as the \textit{size-depth trade-off}.

Before proceeding to the last step of modification of the MDP families to ensure the property $c)$, it is worth quickly confirming that the other global requirements, $ii)$ and $iii)$, are maintained.

\subsection{Quantum learning using $M_2$}

First, we need to confirm that the oraculization process can be applied, and that it helps the agent solve the MDP optimally. The constructed MDP is not $\eta-$episodic, as the longest path leading to a reset is of length $n\times l$, and the shortest is of length $n$.

Whether having differing (upper bounded) cycle lengths is an issue depends on the exact specification of the memory purging mechanism of the environment, and the specification of the scavenging mechanism of the agent.

{However, since we are dealing with environments which are representable by an MDP, the environmental memory need only store only the most recent state, and the current action, and all other parts of the history may be continuously purged. }

{Further, since the environment is episodic, in the sense that the ``query'' action guarantees that the state occurring two steps in the future does not depend on the current state (as the query action reset the MDP to the root level), we are guaranteed that the scavenged system that the agent holds once it collects the root state will suffice to end an episode (and the overall state is pure, provided the agent submitted a pure state input \footnote{Note, the overall input comprises all the input actions, and while they may be entangled, the overall $n$-action state is pure.}).}

The overall MDP-specified environment, can be constructed using only commuting Hermitian operations, thus in a self-reversible fashion\footnote{Note that a unitary $U$ is self-inverse if and only if it is Hermitian, and a product of two Hermitian operators is Hermitian if they commute.}.
Given scavenging and hijacking capacities (at periods of $n\times l$), this suffices for the construction of $E_q$. 
As mentioned earlier in this Appendix, given that our MDP environments were constructed starting from unitary oracles, the oraculization can be further simplified.

The constructed oracle $E_q$ is a simple extension of the quantum $RFS$ oracle, where the $g$ label is used to indicate the input length, and all other labels are ignored. The original RFS oracle can be constructed in a black-box fashion from $E_q$, hence, there exists a quantum agent, which can simulate the access to the RFS oracle, and, consequently, learn the only rewarding action sequence specified by $s(\emptyset)$ of the underlying RFS problem using the algorithm of \cite{2008_Hallgren}.
Notably, the quantum agent can do this in time $O(n\times l \times \kappa_{RFS})$ where $\kappa_{RFS}$ is the quantum query complexity of the underlying RFS problem. This proves the first technical result given formally with the following lemma.
\LE
The $l\times n-$episodic MDP $M_2$ allows oraculization, and a quantum learning agent can learn the optimal policy in expected time $O(poly( \kappa_{RFS} )),$ where $\kappa_{RFS}$ is the quantum query complexity of the underlying generalized RFS problem.
\EL

\subsection{Hardness of classical learning in $M_2$}

Finally, we need to see the hardness of classical learning in the same environment. 
(At this point, it may be worth while to remind the reader that $l$ is typically chosen to be $O(\log(n))$ to realize instances with the most obvious separation, thus $n \times l \in O(poly(n))$.)

The proof technique we use is that of simulation, which is closely related to standard reduction methods. We will provide an interface (also called a reduction), which simulates the MDP $M_2$ from the RFS oracle, in a black-box fashion. The interface is a poly-time algorithm, making the RFS oracle look like the MDP $M_2$, using only the allowed interfaces. 
Since we are proving classical hardness, we can safely use the copying of the inputs, although even that can be achieved reversibly, hence, remaining compatible with quantum interrogation as well.

 The interface is defined such that it simply returns the current input action histories, up to the final length $l \times n$. It observes the sequence, identifies the actual legal query specified by a terminating action ``q'' , and if found, it inputs it into the RFS oracle, returning the result, and reward if the RFS oracle's outputs the value $b_{\emptyset}$.\footnote{Note, this only occurs if the input was $s(\emptyset)$.}  Otherwise it returns the $\bot$ state.
 Note, rewards are issued only if the query corresponds to the lowest-level query, that is to the guessing of $s(\emptyset)$.
 Now, if there is a learning agent/algorithm that learns the MDP $M_2$ in time $n\times l \times T,$ then the composition of this agent/algorithm with the above specified interface is a classical algorithm that solves the RFS problem in time $T$.  Since there is a super-polynomial separation between the minimal possible $T$ and $\kappa_{RFS},$ there must be a  super-polynomial separation between the quantum, and any classical agent in learning efficiency.
In other words, any $(poly(n)^{-1}, poly(n)^{-1})-$efficient classical learning agent requires at least $n^{-\textup{log}(n)}$ steps, whereas the quantum agent attains the same efficiency (in fact, $(1, 1-O(exp(-n))$ efficiency) in $n^{O(1)}$ steps.

To elaborate on the error, and failure probability, note that only the single $n-$bit sequence of $s(\emptyset)$ leads to the reward. If both the error and failure probability are at most polynomially small, polynomial repetition of the process will identify $s(\emptyset)$ with constant probability (and by further repetition, exponentially small failure probability). Since the rewarding path can be checked deterministically, if there exists a classical agent $A$ with only polynomially small error and failure terms, then there exists a classical agent $A'$ with exponentially small failure and zero error terms, which achieves this performance in at most a polynomially larger number of interaction steps relative to $A$, via simple iteration by the Chernoff bound. We have that
\LE \label{RedM2}
No classical learning agent can achieve $(poly(n)^{-1}, poly(n)^{-1})-$efficiency in $poly(n)$ number of interaction steps in $M_2$.
\EL
The two lemmas together imply our first theorem.
 \TH
 Environments specified by MDP $M_2$ allow a super-polynomial separation between classical and quantum $(\epsilon, \delta)-$efficient learning agents, as long as $\epsilon, \delta$ are not super-polynomially decaying. In particular, the separation holds for constant error and failure parameters.  
\HT

\subsection{{Augmenting the MDP with a stochastic action}}
\label{sec:rgmove}

{The MDPs constructed thus far still do not satisfy all the criteria we had outlined earlier in Section \ref{sec:main}: in particular, the characteristic $c)$ is not fulfilled. }
{Recall that this characteristic of the MDP, amongst other consequences, requires a stochasticity of the transition function, on the relevant/rewarding part of the environmental space. }

The property $c)$ asserts that the optimal behavior must explicitly depend on the state label, and not just on the interaction step counter. As we explained earlier, full determinism of an MDP necessarily violates $c)$. Hence, we need to modify the MDP such that the transition function becomes stochastic. Further, stochasticity must appear in the part of the space that must be visited by optimal agents\footnote{{Without demanding this, we could easily introduce stochastic regions to an MDP which is never visited by the agent, which is a trivial, yet uninteresting solution.}}. This can be achieved by a relatively simple trick which ensures the stochasticity, while still allowing the corresponding oraculization $E_q$ to be realized.

\begin{figure}
 \includegraphics[width=0.4\textwidth, trim=15cm 13.5cm 13cm 5cm,clip=true]{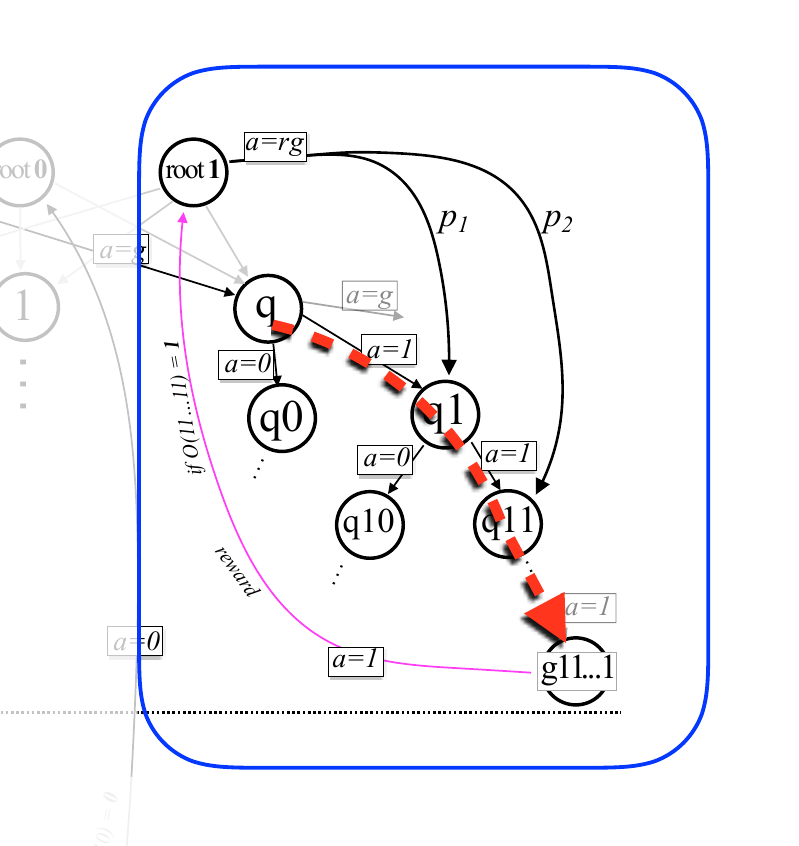}
\caption{\label{MDP3} Illustration of the part of $M_3$ that differs from $M_2$.
The red arrow illustrates the rewarding path. The action $rg$ causes a random transition (with some non-extremal probabilities $0<p_{i}<1,\ i=1,2$ ) to a state in the first half of the rewarding path, from any root element (just root $1$ in illustration).
 }
 \end{figure}

This is realized with the final set of modifications yielding the final MDP $M_3$ which we describe next.

Recall for the moment the structure of the MDP $M_2$ (in Fig \ref{MDP2}), where all the paths are of  length $n\times l$,\footnote{This was enforced by introducing additional ancillary paths which lead to no-rewards, which ensures strict episodicity.} and where all rewarding action paths contain $s(\emptyset)$ as the prefix.\footnote{Recall, in the non-extended MDP, performing the $n$ actions on the rewarding path leads to a re-set. In the strictly episodic variant, the agent still has to perform $n\times(l-1)$ moves before the episode ends, but the choice of moves after the rewarding path does not matter, as all are rewarded.}
We will refer to this unique sequence of percept-actions as \textit{the rewarding prefix}.
Next, for simplicity assume that the rewarding prefix length $n$ is even, and consider the first half of the rewarding prefix -- \textit{the rewarding half-prefix}, which is of length $n/2$. 
We will further expand the action space of the agent, by adding a $rg$ (``random \& guess'') option, which can be advantageously invoked only when paired with the very first bit\footnote{If the agent performs this action at any other position, it is led to a non-rewarding tail of the MDP.}, that is, when the agent is at a root state of the MDP. For intuition purposes, we imagine that the agent separately announces guess options, however, recall that they are actually combined with the bit-specification to ensure uniform path length.
$M_3$ is defined such that, if the agent invokes the $rg$ option at a root state, the subsequent environmental state is randomly chosen from any of the states {appearing in} the half-prefix. 

Note that the optimal course of action for any agent is to \textit{always} invoke the random option, as it, on average, shortens the rewarding sequence by $n/4$. 
However, it is not known beforehand to which position in the rewarding half-path the transition will lead, meaning the optimal agent cannot simply blindly output the rewarding sequence. It is thus forced to learn the full optimal policy specification for the first half of the path. 
Note that the agent, irrespective of the position in the half-path the jump lead to, still needs to exactly reproduce at least the second half of the rewarding sequence.

At this point, we can describe the quantum agent which learn in the MDP $M_3$. We call this agent $A_3$, and we will prove its efficiency shortly. Its behavior has three stages.
In the {first stage} the agent uses oraculization techniques to effectuate/simulate access to the RFS-type oracle, using which it can effectively run the quantum algorithm for RFS, in order to learn $s(\emptyset)$.
In the {second stage}, the agent uses classical access, and blindly executes actions as {prescribed} by the string $s(\emptyset)$ step-by-step, and collect the labels of the states of the MDP along the rewarding path. Note, these labels were not available in the oraculized instantiation (in fact, one of the key steps of the oraculization procedure is the \textit{uncomputation}, or \textit{deletion} of the state labels).
In the final, {third stage}, having learned the correct state-action association along the rewarding path, the agent invokes the randomized short-cut option:  first use the $rg$ action to jump through the first $r$ steps along the rewarding path (for some random $0 \leq r \leq n/2$), then figure out $r$ by looking at the label of the current state of the MDP, then input the remaining $n-r$ bits of $s(\emptyset)$, and obtain a reward, in minimal number of steps. 

For completeness we mention that the performance of the agent is technically evaluated via a tester, which if not chosen carefully could break any entanglement between the agent and environment (see \cite{2015b_Dunjko, 2016_Dunjko} for details). In the quantum stages, the interaction is thus left untested, and then fully classically tested in the classical phases, as was done in \cite{2016_Dunjko}.

{To obtain the central result of this paper, we still need to tie up two loose ends which were comparatively simple for the cases of environments specified by the deterministic MDP $M_2$.
{First, we must make sure a quantum agent can learn in $M_3$ in a polynomial number of interaction steps. To prove this, we need to ensure that the oraculization methods can still be applied in the case of the environments which encode $M_3$ -- note, the transitions of optimal behavior now includes stochastic transitions. } {Although here we consider only the environments specified by $M_3$, the arguments we will use are more broadly applicable.}

Finally, we must ensure that the learning in the constructed MDP $M_3$ is still hard for the classical agent. This is not trivial. Recall, the classical hardness of learning in MDP $M_2$ is easily implied from the classical hardness of the RFS problem. However MDP $M_3,$ which is modified from $M_2$, allows options for the classical agent which do not trivially map to the underlying RFS task.
}

\subsection{Efficient quantum learning in $M_3$ (and other stochastic MDPs)}
\label{upperB}
We first focus on the particular case of the MDPs of the type $M_3$, to show they still allow a consistent oraculization approach. 
Note, intuitively, there could be a problem in these types of environments because the $rg$ move is stochastic and irreversible. Consequently, it is not clear how the quantum environment implements this action. 
However, in our case, this actually causes no issues as the quantum agent $A_3$ never needs to perform quantum superpositions and $rg$ moves simultaneously. Whenever it does an $rg$ move, it is a classical move (not a superposition).

The above observation actually holds more generally. 
Given an MDP $M$, and a subset of actions $A' \subseteq A$, with $M_{|A'}$ we denote the sub-MDP effectuated if the agent is restricted to using only the actions in $A'$.
In the case of $M_3$ we have that ${M_{3}}_{| A \setminus \{ rg\}} = M_2$. In other words, an agent which never uses the randomized guess is effectively in an environment $M_2$, which is deterministic.
But in this case, all the oraculization constructions introduced for deterministic cases \cite{2016_Dunjko} can again be applied. 
This we formulate as the following broadly phrased Lemma.

\LE\label{QuantumUpperBound}
\label{QuantumUB}
Let  $M$ be an MDP specifying a controllable task environment (\textit{i.e.}, a task environment allowing appropriate quantum access). Further, let $M$ be such that it has a \textbf{deterministic} sub-MDP $M'$ with the following two properties: $i)$ having access to a (representation of a) policy which is ($\epsilon-$)optimal for $M'$, allows the efficient learning in $M$, and $ii)$ there exists a quantum learning agent $A'$ which can learn in $M'$ efficiently, using standard (or simplified) oraculization methods . Then there exists a quantum learning agent $A$ which learns efficiently in $M$. 
\EL
\begin{proof}
The  quantum agent $A$, which efficiently learns in $M$, first uses a quantum learning agent $A'$ to learn the $(\epsilon-)$optimal policy for $M'.$ Such an agent $A'$ exists by criterion $ii)$. This trained agent $A'$, or rather, the policy it represents, is the desired representation appearing in criterion $i)$. By the same criterion, there exists a behavior which leads to efficient learning in $M$, and $A$ behaves accordingly. 
\end{proof}
{
To make the broad lemma above more concrete, we can focus on the specific environments we consider in this paper. The sub-MDP $M'$ of $M_3$ in question is exactly $M_2$. An agent $A'$ trained to perform with maximal efficiency in $M_2$ also knows how to navigate the rewarding prefix $M_3$ (and also ``knows'' the secret string $s(\emptyset)$ of the underlying RFS problem).
An agent $A$, which has access to (a simulation of) $A'$ can simply invoke the randomized option $rg$, and from that point on froward the percepts and actions between the environment and the simulation $A'$ which will perform optimally on any part of the rewarding prefix. Such $A$ is an optimal agent. 
Thus, we only need a quantum agent $A'$ to learn how to solve $M_2$, understood as a sub-MDP of $M_3$ realized by restricting the actions in $M_3$ (concretely, by not allowing the option $rg$).
This agent is then ``called'' by $A_3$ to achieve optimal performance in $M_3$, by employing the option $rg$ as often as possible. }

 It should be highlighted that if the agent applies the constructions, and inputs the action leading beyond the deterministic sub-MDP (such as the $rg$ action), the procedure in general breaks: there are no guarantees in the construction on what happens in this case. However, we can always define agents which do not employ such behavior.

We note that certain types of randomized environments, where stochasticity cannot be at all influenced by the agent, can also be beneficially quantized -- such constructions may allow further generalizations of $M_3,$ where random moves occur no matter what the agent does. We leave this for future work. 
Lemma \ref{QuantumUpperBound} established a relatively generic statement, which in combination with the constructions of previous sections shows how to find environments which have interesting desired properties $a) - c)$, while still allowing efficient quantum learning agents. 
{However, as mentioned earlier, whether or not the realized MDPs are still difficult for classical agents depends on the details of the constructions. In the last part of the paper, we prove the classical hardness of learning specifically in $M_3$, thereby proving the existence of environments with a super-polynomial separation for quantum RL.}

\subsection{Hardness of classical learning using $M_3$}
While the classical hardness of learning of $M_2$ was easily implied by the classical hardness of the RFS problem, the modification added to $M_3$ gives the classical agent an option which was not available before. It is easy to see the agent can use this to obtain clearly useful information about the winning path. By using the random jumps, since the state labels specify the correct rewarding path up to the 
state reached by the random transition, the classical agent can, for instance, in $O(poly(n))$ steps the classical agent end up in the very last state of the half-path. This reveals all the moves leading to this point. While learning the rewarding path can be made more convoluted for the agent by permuting the labels (we only need unique labels for the purpose of observability of the MDP), the agent can still build up the connectivity map on the path in $O(n^2)$ queries.

However, as it turns out, even without this modification, we can still prove classical hardness of learning in $M_3$.
We first show an intuitive fact that solving a randomly chosen RFS problem where the prefix is additionally leaked cannot be easier than solving an RFS problem of half the size. 
{This will allow us to prove the main results using one last simple lemma showing that learning in $M_3$ is harder than learning in $M_2$ with additional access to the half-prefix.}

{For technical reasons, we will only consider RFS problems where the function $f$ is the standard inner product. This somewhat restricts the class of all MDPs our results could be applied to as, in principle any generalized RFS (as defined in \cite{2008_Hallgren}) could yield a distinct class of MDPs, but it drastically simplifies the arguments needed for for Lemma \ref{exts}. We discuss the possible generalizations in the Discussion section.}

We begin by first clarify the notation and define a few useful concepts..

First, without loss of generality we assume all the labels of RFS problems are strings of $n$ bits.
Next, with  $RFS(n)$ we denote the family of all such RFS problems with $n-$bit labels. Note, an instance of $RFS(n)$ is fully specified by the function $s$.

Any algorithm which which uses queries to an unknown $RFS(n)$ oracle (specified by the function $s$) to simulate access to another oracle of a problem in $RFS(2n)$ (with some other, higher dimensional secret function $s'$) with the property that $s(\emptyset)$ is a suffix of $s'(\emptyset)$ is called \textit{a lifting of $RFS(n)$,} and the realized larger instance RFS problem $RFS(2n),$ is called \textit{the lifted RFS problem}.

The basic idea here is to show that smaller RFS instances can be black-box embedded in larger instances. This will imply that if one can solve a larger instance if information about the embedding part is leaked, then one can also solve the input and unknown smaller instance, thereby proving the hardness even under leaking of information. 
Any such construction which constructs an instance of $RFS(2n)$ compatible with a given $RFS(n)$ we call a \textbf{lifting construction}. Next, we provide a particular simple lifting construction, which is suitable for standard RFS problems, and which is based on concatenation. 

\paragraph{Concatenated lifting construction} 

Given access to an oracle for one $RFS(n)$, we first choose a full specification of another RFS problem of the same size. Let $s$ be the (unknown) secret function of the first RFS, and $p$ be the (fully known) function of the chosen ancillary RFS. Given a $2n$-bit string $x$, with $\textup{pref}(x)$ we mean the first $n-$bits of $x$, and with $\textup{suff}(x)$ we denote the $n-$bit suffix. Thus $x= \textup{pref}(x)\textup{suff}(x) = \textup{pref}(x)\circ \textup{suff}(x)$, where $\circ$ denotes concatenation (we may omit this symbol when there is no danger of confusion).

The secret function $s'$ of the ``glued'' $RFS(2n)$ is given as follows 
\begin{widetext}
 \EQ{
s'(x_1, x_2,\ldots x_k) = 
p(\textup{pref}(x_1),\textup{pref}(x_2),\ldots,\textup{pref}(x_k)) \circ s(\textup{suff}(x_1),\textup{suff}(x_2),\ldots,\textup{suff}(x_k)).}
\end{widetext}
This construction works naturally with the inner product as $f$, as $f(x,y) = f(\textup{pref}(x),\textup{pref}(y)) \oplus f(\textup{suff}(x),\textup{suff}(y))$.

In other words, we are dealing with two $RFS(n)$ effectively in parallel, with minimal coupling between the subspaces.
It is clear this is a valid construction of $2n-$bit RFS instances. 

\paragraph{Uniform lifting} {The concatenated construction lifts an $n$-bit instance to a $2n$-bit instance of RFS, however, not all $2n$-bit instances of RFS can be realized in this way.
Since we wish to show that learning in $M_3$ is hard in the average case, we also need to show that the capacity to solve an average $2n$-bit RFS with access to the leaked part of the winning sequence is hard. To be able to prove this using the lifting arguments, it is necessary to provide a lifting construction which can realize any $2n$-bit RFS instance, given an appropriate $n-$sized instance.}
 {The instance we capture with the bare concatenated constructions are only those with such a product form, and of course not all instance factorize in this manner}. 
 A lifting procedure which allows the construction of any $RFS(2n)$ compatible with a given $RFS(n)$ we call a \textbf{uniform lifting procedure}.
Here, we show a method on how the lifting construction can be altered to simulate access to another oracle in $RFS(2n)$, such that the modified effective function $s''$ deviates from $s'$ in the value it attains at any chosen path-label by any chosen string of bits.
More specifically, if we chose a path-label (element of domain of $s'$ and $s''$) $\mathbf{x}$, and a 2n-bit string $d$, we construct an oracle specified by $s''$ such that $s'(\mathbf{x}) \oplus_{pw} s''(\mathbf{x}) = d$, where $\oplus_{pw}$ is point-wise mod-2 addition. This is done in a way that ensures that the root values remain compatible with the initial $RFS(n)$ in the suffix.
Moreover, we show that such modifications can be applied sequentially in any order any number of times.
Since we can do this for any chosen vector $d$, and for any set of chosen domain elements $\mathbf{x},$  this means we can realize any element of $RFS(2n)$ compatible with the given, unknown instance of $RFS(n)$. The provided construction uniformly covers the space of lifted $RFS(2n)$ problems, as any can be reached starting from any compatible smaller-instance RFS.

Our basic idea is thus to ``twirl'' the input output relations, on elements of the domain of $s'$ in a black-box fashion: each $2n$-bit secret is randomized, by bitwise xor-ing it with a $2n$-bit shift, which is chosen independently at random. This new $2n$-bit RFS oracle can be simulated using the original $n$-bit RFS oracles, due to bilinearity of $f$. In more detail, we need to show how a ``twirl'' can be effectuated in the actions of the oracle. 

Consider any input $\mathbf{x} = (x_1 \ldots x_k )$ for $2< k <l$, and let $s'(\mathbf{x} ) = z,$ where $z$ is a $2n$-bit string.
Note that the function $s'$ evaluated on $(x_1, \ldots x_k )$ appears in two capacities in an RFS problem. One is a \textit{query-validity criterion} in a call to the oracle of the form $(x_1,\ldots x_k , a),$  where a is a $2n$-bit string.
More precisely we have that
$O_{2n}(x_1,\ldots x_k , a_k) = f(s'(x_1,\ldots x_{k-1} ), x_k) $ if $s'(x_1, \ldots x_k) = a)$, thus we need to 
change the evaluation of the criterion  $s'(x_1, \ldots x_k) = a)$.
The second, \textit{inner-product occurrence} of the evaluation of $s'$ couples the levels of the RFS-tree, where we have that 
$O_{2n}(x_1, \ldots, x_k, x_{k+1}, a) =f(s'(x_1, \ldots, x_k), x_{k+1})$ if $s'(x_1, \ldots x_{k+1}) = a$.

Note that we only wish to change the value $s$ attains on the one query sequence $\mathbf{x}$.
Thus, the simulator, interfacing with the oracle we have, and effectuating an oracle with hidden function $s'',$ with altered values on $\mathbf{x}$ need only act non-trivially when queries are of one of the two forms above.
If it identifies we are in the setting of a query-validity criterion, we wish we could effectuate the test
\EQ{
s'(x_1, \ldots x_k) \oplus_{pw} z = a.
}
However, due to the concatenated construction, this simplifies to
\begin{widetext}
\EQ{
s'(x_1, \ldots x_k) \oplus_{pw} z = a \Leftrightarrow p(\textup{pref}(x_1), \ldots \textup{pref}(x_k)) s(\textup{suff}(x_1), \ldots \textup{suff}(x_k)) = a  \oplus_{pw} z  \Leftrightarrow \\
  p(\textup{pref}(x_1), \ldots, \textup{pref}(x_k)) =  \textup{pref}(a  \oplus_{pw} z)\  \&\&\   s(\textup{suff}(x_1), \ldots, \textup{suff}(x_k))  =  \textup{suff}(a  \oplus_{pw} z).
}
\end{widetext}
Note that the first criterion above we can easily evaluate as it only depends on the known inputs, and a chosen instance of $RFS(n)$. The second, however, can also be checked by querying the other, unknown $RFS(n)$ oracle by inputting only suffixes of the query, and xor-ing the last string with $\textup{suff}(a)$.
In other words, to decide a rejection of input, the simulator checks the two criteria, one of which requires one call to the unknown oracle of $RFS(n)$.

What remains is understanding what the simulator needs to do in the case of an inner product occurrence. We proceed analogously.
We wish to effectuate
$O_{2n}(x_1, \ldots, x_k, x_{k+1}, a) =f(s'(x_1, \ldots, x_k) \oplus d, x_{k+1})$ if $s'(x_1, \ldots x_{k+1}) = a$.
Note that the validity criterion requires no modification. To evaluate $f(s'(x_1, \ldots, x_k) \oplus_{pw} d, x_{k+1})$ we have that, by distributivity:
\EQ{
f(s'(x_1, \ldots, x_k) \oplus_{pw} d, x_{k+1}) =\nonumber\\
 f(s'(x_1, \ldots, x_k), x_{k+1}) \oplus f(d, x_{k+1}).
}
In other words, the simulator need only xor the output (if the instance passed the validity criterion) with the value $f(d, x_{k+1})$, which it can easily do.

Thus the simulator can effectuate an $RFS(2n)$ with the underlying labeling function $s''$ using only access to an (effective) oracle realizing the functions $s$ and $p$. Note also that $s''$ differs from $s'$ in any chosen domain value, where it can attain any possible label. However, to maintain compatibility of the output, we should not alter the values at the root, {even though such alterations could be decoded as well, if needed}.

Note that the alterations of the type above can easily be ``stacked'', or composed as the occurrences are independent.
This means that any labeling function can be obtained, which allows us to achieve uniform sampling over $RFS(2n).$
This implies the following lemma:
{
\LE\label{exts}
There exists a uniform lifting procedure which can, given black-box access to an instance $R$ of $RFS(n)$ with secret string $s(\emptyset)$, generate all instances of $RFS(2n)$ such that the secret string $s'(\emptyset)$ of the generated sequences contains $s(\emptyset)$ as the suffix.
\EL
In particular, this implies that given uniform distribution over black-box input instances, we can construct a uniformly distributed instance from the set $RFS(2n)$.}

Next, we show that the existence of such a uniform lifting construction implies that solving $RFS(n)$ is still hard for the classical agent, even in the full information about the prefix of the secret string $s(\emptyset)$ is leaked, also in the average case.

\LE\label{LeakHard}
If there exists a construction which uniformly covers the space of  lifted $RFS(2n)$ problems, then solving a random instance of $RFS(2n)$ given additional information about the half-prefix cannot be done more efficiently than solving a random (or a worst case) instance of $RFS(n)$. 
\EL
\begin{proof}
Suppose there is a construction as in the statement of the theorem, and an algorithm $P$ which solves a random $RFS(2n)$ in time $T$, given side information about the half-prefix.
Then given a random instance of $RFS(n)$ we construct a uniform sample from $RFS(2n)$ compatible with the smaller instance using the construction above.
Denote the secret string of the smaller instance $s(\emptyset)$, where the secretes of larger instances are denoted with  $s'(\emptyset)$.
 Note, the construction need not be efficient, only computable, as we only care about query complexity. The construction will realize some $RFS(2n)$ with a known prefix. Since the small instance was uniformly sampled and since the construction generates an uniform sample from the compatible class, the output instance of the $RFS(2n)$ also from a uniform distribution. 
This valid input instance is, along with the side information about the prefix, fed into $P$, and the output is collected. The suffix of the output is, by construction, the solution of the smaller instance. 

Note, the only compatibility constraint we maintain is the values of the suffix of $s(\emptyset)$ of the larger instance, and this is the only property we maintain from the small instance. However, even this can be modified by a chosen deviation string $d$ (see details of construction later) -- i.e. we can construct an instance of $RFS(2n)$ with the only constraint that the suffix of $s'(\emptyset)$ is equal to $s(\emptyset) \oplus_{pw} d$, for any $d$.
That is we can generate any instance of $RFS(2n),$ and by obtaining the solution to this instance, recover the solution to the smaller instance. This means that we can also drop the assumption on the uniformity of the small instance, and the claim holds for the worst case as well.
\end{proof}
{
Now we can prove the classical hardness of learning in $M_3$.}

\LE\label{MainLem}
No classical learning agent can achieve $(poly(n)^{-1}, poly(n)^{-1})-$efficiency in $poly(n)$ number of interaction steps in $M_3$.
\EL

\begin{proof}
First, note that if learning in $M_2$, in the setting where the agent is given prior information about the half-prefix of the winning path is hard (i.e. $(poly(n)^{-1}, poly(n)^{-1})-$efficiency is not attainable in $poly(n)$ time for any classical agent), then learning in $M_3$ is hard as well. To see this note that knowing the prefix of $M_2$ allows the agent to black-box simulate the corresponding $M_3$ from $M_2$, by randomly choosing a position in the winning prefix in $M_2,$ and deterministically moving to that point.
Hence, the existence of an efficient agent for $M_3$ with leak implies the existence of an  an efficient agent for $M_2$ with information about the prefix. 

What remains to be seen is that learning in $M_2$ is hard even when the prefix is leaked.
 Lemma \ref{LeakHard}, along with (the existence of) the uniform lifting construction implies that the solving of $RFS(n),$ when half of the secret string is leaked is still classically hard.
 Then, by the same simulation constructions we have given in the proof of Lemma \ref{RedM2} (which precede the statement of the lemma), it follows that an efficient agent learning $M_2$ given half of the winning prefix implies that the corresponding ``leaky'' $RFS(n)$ can be solved as well, which is not possible.
 This concludes the proof.
\end{proof}

\subsection{Quantum advantage for learning in $M_3$}

{Combining the above result on classical hardness of learning in $M_3$, with the result in Section \ref{upperB} showing how efficient quantum learning agents can be constructed for $M_3$, we have our main theorem:}

  \TH
 Environments specified by MDP $M_3$ allow a super-polynomial separation between classical and quantum $(\epsilon, \delta)-$efficient learning agents, as long as $\epsilon, \delta$ are not super-polynomially decaying. In particular, the separation holds for constant error and failure parameters.  Finally $M_3$ satisfies all three criteria $a)-c)$ for MDPs with generic properties.
\HT
\begin{proof}
The theorem follows from the proof of classical hardness of $M_3$ in Lemma \ref{MainLem}, and the existence of a quantum agent which can efficiently learn in $M_3$, which is implied by Lemma \ref{QuantumUpperBound}. In essence, the latter relied on the possibility of applying oraculization techniques to instantiate $E_q$ which is, on the relevant subspace, equal to the standard $RFS$ oracle, the existence of the efficient quantum algorithm for RFS, which reveals the entire string $s(\emptyset)$.
The fact that $M_3$ satisfies all three criteria $a)-c)$ holds by construction. \end{proof}
 
 \subsection{{RFS as an example of hierarchical learning}}
 \label{self:sim}
{Textbook RL tasks deal with MDP environments, and problems like optimizing return in the infinite or finite horizons, 
on a more abstract level. However, aside from learning how to solve the local problem, intelligent agents are expected to also exploit previous experiences in future situations. One aspect of this is a form of generalization, where agents trained to solve some problem $E$ should have an advantage in a similar problem $E',$ relative to an untrained agent. While the possibility for such generalization critically depends on the notion of ``similarity'', one way of understanding a key feature of generalization is in a type of information carry-over: optimizing in $E$ provides the agent with useful information for solving $E'$. 

The proposals which achieve a quantum speed-up in \cite{2016_Dunjko} and also in this work, can be seen as examples of this paradigm: the quantum agent accesses a (simplified quantum) environment $E_q$ to learn a key needed for optimal performance in the full environment. One can generalize this, and imagine a cascade of problems $\{ A^{(k)}_k\}$, where solving of any task $k$ requires (or is simply significantly mitigated) the solution of $k-1$. In natural learning settings, an example of this is for instance hierarchical skill learning \cite{2016_Hangl}, where an agent is trained in simpler skills, which are later used as primitives for more complex tasks -- beyond robotics, this is also how biological agents learn.}

{The standard RFS problem itself has a flavor of such hierarchical learning: it is a recursive problem, where the queries to lower-level instances are locked by the solutions of higher-level instances (in the RFS tree)\footnote{{Specifically, each RFS tree of depth $n$, with $n$-bit labels encodes smaller-instance RFS problems. The smaller instance-tree is a sub-tree of the overall graph. A corresponding sub-tree can be realized as follows. First, the last layer is dropped. Second, starting from the root, we choose a bit, and identify all children whose label does not end with the chosen bit -- those nodes, along with their sub-trees, are dropped. This is iterated on each remaining node, sequentially through the layers.  What remains is tree for RFS problem over $(n-1)$-bit labels.}}.
At the same time, unlocking the secret of a higher-level does not actually help in the unlocking of the secrets of the lower-level instances.
} 
{
Note, if it were the case that the secrets between layers share known correlations, the problem would likely have a more efficient classical solution, which utilizes the learned secret strings. On the other hand, the quantum algorithm does rely on certain consistency criteria (the secrets of the higher level unlock the oracles of the lower), but not on any correlations between secrets of the layers.

In a sense, the RFS problem may constitute an example of hierarchical learning suitable for quantum learners, but not for classical learners. We plan on exploring this possibility, and its relevance for the theory of (quantum) RL further in subsequent works.
}
{With this, we move onto the next example, where the underlying oracle does not have a recursive structure, and the ``secret'' is hidden in the global properties of the underlying MDP, based on the Simon's problem.}

\end{document}